\documentclass[aip,reprint,floatfix]{revtex4-1}
\usepackage{graphicx}
\usepackage{dcolumn}
\usepackage{bm}
\usepackage{amsmath}
\usepackage{xcolor}

\usepackage[utf8]{inputenc}
\usepackage[T1]{fontenc}
\usepackage{mathptmx}
\usepackage{etoolbox}

\def\beq{\begin{equation}}
\def\eneq{\end{equation}}

\def\e{{\hbox{e}}}

\def\uq*o{{\left<U^{o}_{q^*}U^{o}_{-{q^*}}\right>}}

\def\l{\left}
\def\r{\right}

{\large }

\def\lb{\label}

\makeatletter
\def\@email#1#2{%
 \endgroup
 \patchcmd{\titleblock@produce}
  {\frontmatter@RRAPformat}
  {\frontmatter@RRAPformat{\produce@RRAP{*#1\href{mailto:#2}{#2}}}\frontmatter@RRAPformat}
  {}{}
}%
\makeatother

\begin{document}

\preprint{AIP/123-QED}

\title{Active fractal networks with stochastic force monopoles and force dipoles: Application to subdiffusion of chromosomal loci}

\author{Sadhana Singh}
\affiliation{Avram and Stella Goldstein-Goren Department of Biotechnology Engineering, Ben-Gurion University of The Negev, Beer Sheva 84105, Israel}
\author{Rony Granek}
\affiliation{Avram and Stella Goldstein-Goren Department of Biotechnology Engineering, Ben-Gurion University of The Negev, Beer Sheva 84105, Israel}
\affiliation{Department of Physics, and Ilse Katz Institute for Nanoscale Science and Technology, Ben-Gurion University of The Negev, Beer Sheva 84105, Israel}
\email{rgranek@bgu.ac.il}

\date{\today}

\begin{abstract}\label{sec:abs} 
Motivated by the well-known fractal packing of chromatin, we study the Rouse-type dynamics of elastic fractal networks with embedded, stochastically driven, active force monopoles and force dipoles that are temporally correlated. We compute, analytically -- using a general theoretical framework -- and {\it via} Langevin dynamics simulations, the mean square displacement (MSD) of a network bead. Following a short-time superdiffusive behavior, force monopoles yield anomalous subdiffusion with an exponent identical to that of the thermal system. In contrast, force dipoles do not induce subdiffusion, and the early superdiffusive MSD crosses over to a relatively small, system-size-independent saturation value. In addition, we find that force dipoles may lead to ``crawling" rotational motion of the whole network, reminiscent of that found for triangular micro-swimmers and consistent with general theories of the rotation of deformable bodies. 
Moreover, force dipoles lead to network collapse beyond a critical force strength, which persists with increasing system size, signifying a true first-order dynamical phase transition. We apply our results to the motion of chromosomal loci in bacteria and yeast cells' chromatin, where anomalous sub-diffusion, MSD$\sim t^{\nu}$ with $\nu\simeq 0.4$, was found in both normal and ATP-depleted cells, albeit with different apparent diffusion coefficients. We show that the combination of thermal, monopolar, and dipolar forces in chromatin is typically dominated by the active monopolar and thermal forces, explaining the observed normal cells {\it vs} the ATP-depleted cells behavior. 
\end{abstract}

\maketitle

\begin{quotation}
The slowed, subdiffusive motion of chromosomal loci within cells has recently become a focal point of research. This phenomenon has been particularly noted in experimental studies involving both normal cells, where active forces are prevalent, and ATP-depleted cells, where forces are purely thermal. Interestingly, in both cases, the observed subdiffusive exponents ($\simeq 0.4$) remain consistent, a finding that current theories struggle to explain. Fractals, intricate self-similar patterns pervasive in nature, have drawn new interest due to their relevance to chromatin structure, which has been firmly established through experiments. Motivated by this fractal nature of chromatin, our theoretical investigation explores the dynamics of a fractal network influenced by active forces—both monopolar and dipolar. We aim to elucidate why chromosomal loci exhibit identical subdiffusive exponents in both normal and ATP-depleted cells.

\end{quotation}

\section{Introduction}
Active matter is attracting increasing interest and spans diverse systems \cite{ramaswamy_2010_anu_rev,active_marchetti_2013,cates_2015_review,active_2015,cates_2016_active_matter,active_fodor_2018}. Much effort has been recently devoted to the phase diagram of active systems, 
with emphasis on ``motility induced phase separation" \cite{cates_2015_review,phase_gonnella_2015,mips_2018,misp_2020,mips_soft_2021,mips_2022}. Living systems, where the chemically stored energy is transformed to mechanical energy, can be regarded as a type of active matter, e.g., moving micro-organisms, {\it in-vivo} intracellular dynamics, or {\it in-vitro} active cytoskeleton dynamics \cite{cells_wilhelm_2008,invivo_robert_2010,cell_toyota_2011,cytoskeleton_mackintosh_2012,cytoplasm_guo_2014,living_matter_fodor2015,active_living_2024}. Earlier studies of such bio-systems aimed to obtain steady-state near-equilibrium dynamics, in particular, its manifestation in the mean square displacement (MSD) of a probe monomer or particle \cite{granek_1999,granek_2000,msd_barkai_2012,msd_ghosh_2022,subdiffusion_garini_2009_prl,lamina_a_garini_2015,viscoelastic_weber_2010,weber_2012}.

In one such recent {\it in-vivo} experiment, the motion of different chromosomal loci in bacteria and yeast cells was probed both in normal cells, in which active ('ac') forces dominate, and in ATP-depleted cells, where fluctuations are governed by thermal ('th') forces \cite{viscoelastic_weber_2010,weber_2012}. It was found that in both cases, chromosomal loci perform anomalous sub-diffusion, MSD$\approx B\,t^{\nu}$ ($\nu<1$), with identical exponents $\nu_{ac}\simeq \nu_{th}\simeq 0.4$ yet different amplitudes $B$ (from which an apparent diffusion coefficient is defined in Ref. \cite{weber_2012} as $B=2d D_{\text{app}}$, where $d$ is the dimensionality); the ATP-depleted cells yield a lower amplitude. Telomeres in chromatin of mammalian cells also exhibit subdiffusion with a similar exponent, $\nu_{ac}\simeq 0.4$ \cite{subdiffusion_garini_2009_prl,lamina_a_garini_2015}. Yet, the reason behind the equality of the two exponents reported in Ref. \cite{weber_2012} is unclear \cite{mackintosh2012active}. While passive (thermal) and active Rouse dynamics of linear chains predict identical subdiffusion exponents, their values are $\nu_{ac}=\nu_{th}=1/2$, \cite{osmanovic_2017,winkler_2020} larger than the observed values. 

To explain these discrepancies, we take advantage of the fractal structure of chromatin (discussed next). Fractals, either disordered or deterministic, are characterized by a few broken dimensions \cite{stauffer_1992,bunde1995branched,ben_avraham_book_2000}: (i) the mass fractal (Hausdorff) dimension $d_{f}$ governing the scaling $M(r)\sim r^{d_{f}}$ of the mass $M(r)$ enclosed in concentric spheres of radius $r$, (ii) the spectral dimension $d_{s}$ governing the scaling $g(\omega)\sim\omega^{d_{s}-1}$ of the vibrational density of states (DOS) $g(\omega)$ with frequency $\omega$ \cite{alexander_1982,alexander_1989}, and (iii) the topological dimension $d_{l}$, that governs the scaling $M(l)\sim l^{d_{l}}$ of the mass $M(l)$ enclosed in concentric ``spheres" of radius $l$ in the topological (or ''manifold'' [''chemical''] space. One may also define, instead of $d_l$, the chemical length (or minimal path) dimension $d_{\text{min}} = d_f/d_l$, which relates the real space distance $r$ between two points on the fractal to the minimal path distance $l$ between these points along the fractal network links, 
$l \sim r^{d_{\text{min}}}$ \cite{bunde_1992,bunde_1997,ben_avraham_book_2000}. To clarify the concept of topological space and $d_l$, consider a linear chain -- defining a 1D topological space -- folded self-similarly in the 3D space without forming crosslinks, thereby setting $d_l=1$ independent of $d_f$ (e.g., $d_f=2$ and $5/3$ for Gaussain and self-avoiding chains, respectively), while crosslinks may increase $d_l$ above 1. Similarly, a 2D sheet may be crumpled self-similarly in the 3D space without internal crosslinks, corresponding to $d_{l}=2$ irrespective of $d_f$. The three broken dimensions, $d_s$, $d_l$, and $d_f$, obey the inequalities: $1\leq d_s \leq d_l\leq d_f \leq d$ where $d$ is the Euclidean embedding space dimension.

In recent years, significant insight has been gained in understanding chromatin structure during the interphase stage using various experimental methods \cite{fractal_einstein_1998,fractal_lebedev_2005,fractal_lebedev_2008,fractal_lieberman_2009,fractal_bancaud_2009,fractal_wax_2002,fractal_chalut_2009,fractal_bancaud_2012,fractal_yi_2015}. These experiments reveal a fractal-like density distribution and estimate the values for the (mass) fractal dimension $d_f$ (see Refs. \cite{fractal_bancaud_2012,iashina_2019_fractal} for summaries of values). In particular, Hi-C experiments \cite{fractal_lieberman_2009}, textural analysis \cite{fractal_einstein_1998}, and neutron scattering \cite{fractal_lebedev_2005,fractal_lebedev_2008} yielded values of $d_f\simeq 2.8-2.9$. The latter values of $d_f$ are nearly those predicted by the fractal globule model \cite{grosberg_1988,grosberg_1993,fgm_2023}, that builds on a linear polymer folded compactly in 3D space, thereby fractal dimensions (see next paragraph) $d_l=d_s=1$ and $d_f=3$. Given the abundance of lamin A proteins making internal crosslinks \cite{lamina_a_garini_2015}, we may expect that both $d_l$ and $d_s$ are somewhat larger than 1.

What are the anticipated effects of active forces acting on chromatin within living cells? These forces can arise from various sources \cite{albert_2002, mackintosh_2008,shivshankar_2017,bruinsma_2014}. Here, we focus on two main categories: force monopoles and force dipoles. Force monopoles are external forces that cannot emerge from within the network, i.e., exerted by the cytoplasm and membrane (in prokaryotic cells) or by the cytoskeleton (in eukaryotic cells) \cite{shivshankar_2017}. On the other hand, force dipoles are internal forces predominantly found in the cytoskeleton, such as the actomyosin system. They might also be present in chromatin, potentially through various enzymes that actively interact with DNA \cite{bruinsma_2014}. Higher-order multipoles, such as force quadrupoles, cannot be excluded completely when considering complicated enzyme or motor protein action but are likely rarer and weaker than the monopoles and dipoles.

Explanations have been proposed for the observed anomalous diffusion exponents in chromatin, $\nu_{ac}\simeq \nu_{th}\simeq 0.4$, yet they do not account for the behaviors of {\it both} normal and ATP-depleted cells on the same footing. One approach, based on a linear chain topology, has been to associate $\nu_{th}<1/2$ with the presence of a viscoelastic solvent, whose frequency-dependent complex modulus behaves as a power-law, $G^*(\omega)\sim (i\omega)^{\alpha}$, 
yielding $\nu_{th}=\alpha/2$ \cite{viscoelastic_weber_2010,viscoelastic_vandebroek2014,visco_vandebroek_2015,viscoelastic_polovnikov2018}. However, a different exponent (also discussed briefly in Appendix \ref{append_sec_viscoelastic} and Sec. II.A.1), $\nu_{ac}=3\alpha/2 -1$, is predicted for an equivalent active system \cite{visco_vandebroek_2015}, contradicting the experimental results. 
Another explanation was based on Rouse dynamics (of a folded linear chain) yielding
$\nu_{th}=2/(2+d_f)$ \cite{fractal_anomalous_2015}. Alternatively, it has been shown that this exponent is always $\nu_{th}=1-d_s/2$, irrespective of the fractal dimension 
\cite{granek_2005,granek_2011,shlomi_2012_pre,shlomi_2012_prl}; hence, $\nu_{th}=1/2$ is recovered for $d_s=1$. These two results for $\nu_{th}$ become identical for Gaussian fractal networks (i.e., in the absence of self-avoidance) for which $d_f=2d_s/(2-d_s)$ \cite{cates1985flory,vilgis1988flory}. This implies that $d_s=6/5$ corresponds to $d_f=3$, leading to $\nu_{th}=0.4$, consistent with the experimental results for ATP-depleted cells. From a different perspective, a single particle model, that uses the generalized Langevin equation with a delta-correlated active force that does not obey the fluctuation-dissipation theorem (FDT), yields $\nu_{ac}=2\nu_{th}-1$ for $\nu_{th}\geq 1/2$ and $\nu_{ac}=0$ for $\nu_{th}<1/2$, such that identical passive and active exponents are obtained only for normal diffusion, $\nu=1$ \cite{granek_2000,caspi_2002,visco_vandebroek_2015,mackintosh2012active}. Active dynamics of critical percolation clusters -- forming disordered fractals -- of a 2D triangular network of springs have been studied previously, yet the individual node MSD was not reported \cite{isostatic_spring_2019}. Here, we aim to generalize such models to arbitrary fractals with potential application to chromatin.

This paper is structured as follows: In Section~\ref{sec_model}, we begin with an overview of the model, the general analytical theory that solves it (detailed in Appendix~\ref{append_sec_analytical}) that applies to any fractal, and the primary results for MSD. Next, in Section~\ref{sec_simulation}, we present simulation results for the Sierpinski gasket, a well-known deterministic fractal, thereby validating our analytical results and discovering new findings for the case of force dipoles. In Section~\ref{sec_chromatin}, we extend the present general theory of active fractal network to chromatin relying on its fractal nature and show general agreement with experimental findings. We conclude in Section~\ref{sec_discussion} by highlighting the role of active forces in active fractal dynamics and the significance of the nature of active forces in chromatin dynamics.

\section{Model and analytical theory}\label{sec_model}
Here, we provide an outline of the analytical theory. The complete derivation is described in \ref{append_sec_analytical}. Following the formalism developed for passive fractals \cite{granek_2005,granek_2011,shlomi_2012_pre,shlomi_2012_prl}, we consider a fractal network of beads connected by harmonic springs where, in addition to the white thermal noise, beads experience stochastic active forces distributed randomly but uniformly over the network. In the case of force monopoles, the force field $\vec{F}({\vec r},t)$ at position $\vec{r}$ is
\beq
\vec{F}({\vec r},t)=\sum_j \vec f_j(t) \delta(\vec r-\vec r_j),
\label{monopole}\eneq
where $\vec{f}_j(t)$ is an active force acting on bead $j$ at position $\vec r_j$. For force dipoles,
\beq
\vec{F}({\vec r},t)=\sum_j \vec f_j(t) \left[\delta(\vec r-\vec r_j)-\delta(\vec r-\vec r_j-\vec\epsilon_{j})\right],
\label{dipole}\eneq
where $\vec{\epsilon}_{j}$ is a randomly chosen vector from $j$ to one of its nearest-neighbor beads ($|\vec{\epsilon}_{j}|=b$), and $\vec f_j$ is taken parallel to $\vec{\epsilon}_{j}$. The limit $\epsilon_{j}\to 0$ should be taken in the continuum limit of point force dipoles. A positive $\vec f_j$ implies an inward (contractile) force dipole, and a negative $\vec f_j$ implies an outward (extensile) one \cite{ramaswamy_2002_prl,ramaswamy_2010_anu_rev,dipoles_intro_de_2015}.

The stochastic active forces are assumed to fluctuate independently in time between $f_0$ (probability $p$) and $0$ (probability $1-p$), following the random telegraph process \cite{gardiner_1991}. 
The auto-correlation function of the forces thus follows
\beq
\langle \vec f _i(t)\cdot \vec f _j(0)\rangle =f_0^2p^2\hat f_i\cdot\hat f_j+f_0^2\delta_{ij}p(1-p)\e^{-t/\tau},
\lb{append_force_corr-1}
\eneq
where $t$ is lag-time and $\tau$ is the force correlation time. We assume that the force variables $f_i$'s, appearing in Eqs.\ (\ref{monopole})-(\ref{dipole}) are {\it not} correlated in space. In addition, their directions are assumed isotropically distributed (see further discussion in Appendix \ref{append_sec_analytical}).

To simplify the analytical treatment, we use the scalar elasticity Hamiltonian \cite{elasticity_nakayama_1994,granek_2005} for the displacements $\vec{u}_i=\vec{r}_i-\vec{r}_{0,i}$ about the equilibrium positions $\{\vec{r}_{0,i}\}$,
\beq
H\l[\{\vec{u}_i\}\r] = \frac{1}{2}m\omega_o^2\sum_{<ij>} \left(\vec{u}_i-\vec{u}_j\right)^2,
\label{hamiltonian}
\eneq
where $\omega_o$ is the spring self-frequency, $m$ is the bead mass, and $<ij>$ stands for nearest-neighbor pairs connected by springs. The Langevin equations of motion in the overdamped limit, assuming local friction (i.e., a Rouse-type model), are
\begin{equation}
\gamma \frac{d\vec u_i(t)}{dt} =m\omega_o^2\sum_{j\in i} \left(\vec{u}_j-\vec{u}_i\right) +\vec{\zeta}_i(t)+\vec f_i(t)
\label{Langevin1}
\end{equation}
($i=1,...,N$), where $j\in i$ stands for beads that are connected (as nearest-neighbors) to $i$, $\vec f_i(t)$ is the active force acting on bead $i$, $\gamma$ is the local friction coefficient, and $\vec{\zeta}_i$ is the thermal white noise obeying FDT, \cite{kubo_1966}
$\langle\vec{\zeta}_i(t)\cdot\vec{\zeta}_j(t^{\prime})\rangle=2d
k_BT\gamma\delta_{i,j}\delta(t-t^{\prime})$, where $T$ is the temperature and $k_B$ is 
the Boltzmann constant. Hence, its variance is
$\langle \zeta_i^2\rangle = 2dk_B T \gamma$, while that of the active noise -- for $t\gg \tau$, where the active noise becomes effectively white -- is $\langle f_i^2\rangle = 2f_0^2\tau p(1-p)$. This allows comparison between the relative strengths of thermal and active forces.

Using the normal modes of the fractal network $\Psi_{\alpha}$ (for details, see Appendix \ref{append_sec_analytical}), whose corresponding eigenfrequencies are $\omega_{\alpha}$, and transforming Eq.~(\ref{Langevin1}) to the normal mode space, we obtain
\begin{equation}
\frac{d\vec u_{\alpha}}{dt} = -\Gamma_{\alpha}u_{\alpha}+\vec{\zeta}_{\alpha}(t)+ \Lambda_{\alpha}\vec{F}_{\alpha}(t) 
\label{Langevin4}
\end{equation}
Here $\vec u_{\alpha}(t)$ is the amplitude of a normal mode $\Psi_{\alpha}$ at time $t$, $\Gamma_{\alpha}=m\omega_{\alpha}^2 \Lambda_{\alpha}$ is the mode relaxation rate, where $\Lambda_{\alpha}=1/\gamma$ is the mode mobility coefficient. $\vec{\zeta}_{\alpha}(t)$ and $\vec{F}_{\alpha}(t)$ are (respectively) the mode-transformed thermal white noise and active noise field; the latter takes the form: (i) For force monopoles 
\beq
\vec{F}_{\alpha}(t)=\sum_j \delta\vec f_j(t) \Psi_{\alpha}(\vec r_j)
\eneq
and (ii) for force dipoles
\beq
\vec{F}_{\alpha}(t)=\sum_j \delta\vec f_j(t) \left[\Psi_{\alpha}(\vec r_j)-\Psi_{\alpha}(\vec r_j+\vec\epsilon_{j})\right],
\eneq
where $\delta\vec{f}_j(t)=\vec{f}_j(t)-f_0p\hat{f}_j$ are the force fluctuations about their mean, and $\Psi_{\alpha}(\vec{r_j})\equiv \Psi_{\alpha,j}$ is the $j$-th entry of normal mode $\alpha$.

Next, we calculate the auto-correlation function of the active force field in mode space (Appendix \ref{append_steady})
\beq
\langle\vec{F}_{\alpha}(t)\vec{F}_{\alpha}(0)\rangle \simeq W_{\alpha}\phi f_0^2p(1-p)\e^{-t/\tau},
\eneq
where $W_{\alpha}=Q\omega^{\xi}$ :  (i) $Q=1$, $\xi=0$, for force monopoles, and (ii) $Q=2 C_0\omega_o^{-2d_s/d_l}$, $\xi=2d_s/d_l$, for force dipoles. Here, $\phi$ is the (force density) fraction of force monopoles or dipoles in the system, and $C_0$ is a numerical constant related to the mean localization properties of the normal modes \cite{alexander_1982,alexander_1989,stauffer_1992,bunde_1992,bunde_1997}.

\subsection{Mean square displacement} Solving the Langevin equation (\ref{Langevin4}) (see Appendix \ref{append_msd} for details), expanding the MSD of an internal network bead, $\langle\Delta\vec{r}(t)^2\rangle\equiv\langle\left(\vec{r}_i(t)-\vec{r}_i(0)\right)^{2}\rangle= \langle\left(\vec{u}_i(t)-\vec{u}_i(0)\right)^{2}\rangle$, in terms of the normal modes, and performing disorder averaging, we obtain the MSD as the sum of a purely thermal contribution, $\langle\Delta\vec{r}(t)^2\rangle_{\text{th}}$, and a purely active contribution, $\langle\Delta\vec{r}(t)^2\rangle_{\text{ac}}$,
\begin{equation}
 \langle\left(\vec{r}(t)-\vec{r}(0)\right)^{2}\rangle=\langle\Delta\vec{r}(t)^2\rangle_{\text{th}}+\langle\Delta\vec{r}(t)^2\rangle_{\text{ac}}
 \end{equation}
 where
\begin{subequations}
\begin{align}
\begin{split}
& \langle\Delta\vec{r}(t)^2\rangle_{\text{th}} = {\frac{1}{N}}\sum_{\alpha} {\frac{2d k_BT}{m \omega_{\alpha}^2}}\left(1-\e^{-\Gamma_{\alpha}t}\right),
\label{MSD_total_th}
 \end{split}\\
\begin{split}
 &\langle\Delta\vec{r}(t)^2\rangle_{\text{ac}} = \frac{1}{N}\sum_{\alpha}\frac{2\phi f_0^2p(1-p) W_{\alpha}\Lambda_{\alpha}^2}{\Gamma_{\alpha}\left(\Gamma_{\alpha}+\tau^{-1}\right)}\\
  & \times\left(1 + {\frac{\tau^{-1}}{\Gamma_{\alpha}-\tau^{-1}}\e^{-\Gamma_{\alpha}t}} -
  {\frac{\Gamma_{\alpha}}{\Gamma_{\alpha}-\tau^{-1}}\e^{-t/\tau}} \right)
\end{split}\label{MSD_total_ac}
\end{align}
\end{subequations}
(where $d$ is the Euclidean dimension). 

Using the vibrational DOS $g(\omega)\sim \omega^{d_s-1}$, we can approximate the sum in Eq.~(\ref{MSD_total_ac}) to an integral over the frequency, 
where the lower and upper integration limits determine (respectively) the shortest, 
$\tau_0=\Gamma(\omega_0)^{-1}=\gamma/m\omega_0^2$, and the longest, 
$\tau_N\sim \Gamma(\omega_{\text{min}})^{-1} \sim \tau_0 N^{2/d_s}$ relaxation times of the system. 
As shown in Refs \cite{granek_2005,granek_2011,shlomi_2012_pre,shlomi_2012_prl}, focussing on the time 
regime $\tau_0\ll t\ll \tau(R_g)$, the thermal MSD exhibits subdiffusion
\beq
\langle\Delta\vec{r}(t)^2\rangle_{\text{th}}=B_{\text{th}} t^{\nu_{\text{th}}},
\eneq
with $\nu_{\text{th}}=1-\frac{d_s}{2}$, provided that $d_s<2$. The prefactor is $B_{\text{th}}\sim dk_BT\gamma^{d_s/2-1}$.

For a large system such that the frequency spectrum is dense, $\langle\Delta\vec{r}(t)^2\rangle_{\text{ac}}$ can also be evaluated analytically. Assuming short times such that $t\ll \tau\ll \tau_N$, we find, for both force monopoles and force dipoles, superdiffusive, ballistic-type behavior,
\beq
\langle\Delta\vec{r}(t)^2\rangle_{\text{ac}}\simeq B_{\text{acs}} t^2\;\;\;   t\ll \tau,
\label{superdiff}\eneq
where the amplitude follows the scaling: (i) for force monopoles, $B_{\text{acs}}\sim \phi f_0^2\tau^{-d_s/2}$ provided that $d_s<2$, and (ii) for force dipoles, given that $d_s/2 + d_s/d_l>1$, we have $B_{\text{acs}}\sim \phi f_0^2\tau^{-1}$. Otherwise, for rare fractal geometries where $d_s/2 + d_s/d_l<1$, we get $B_{\text{acs}}\sim \phi f_0^2\tau^{-d_s/2-d_s/d_l}$.

\subsubsection{Force monopoles} For times much longer than the force correlation time such that $ \tau \ll t\ll \tau_N$, we find for force monopoles
\beq
\langle\Delta\vec{r}(t)^2\rangle_{\text{ac}}= B_{\text{ac}}t^{\nu_{\text{ac}}},
\label{subdiff}\eneq
where $\nu_{\text{ac}}=1-d_s/2$, that is {\it identical} to the thermal subdiffusion exponent. The (purely) active amplitude is $B_{\text{ac}}\sim \phi f_0^2\tau p(1-p)\gamma^{d_s/2-2}$ such that 
\beq
\frac{B_{\text{ac}}}{B_{\text{th}}}= \frac{\phi f_0^2\tau p(1-p)}{dk_BT \gamma}. 
\eneq
Hence, if the active system combines thermal ($T>0$) and active forces (denoted by 'eff' subscript), e.g., as would be expected for normal cells, the MSD is $\langle\Delta\vec{r}(t)^2\rangle= B_{\text{eff}}t^{\nu}$ where $B_{\text{eff}}=B_{\text{ac}}+B_{\text{th}}$, and the amplitude ratio is 
\begin{equation}
\frac{B_{\text{eff}}}{B_{\text{th}}}=1+ \frac{\phi {f_0}^2\tau p(1-p)}{d k_BT\gamma}.
\label{beff}
\end{equation}

In addition, for arrested center-of-mass (CM) translation motion, the MSD saturates at a value $2\langle\vec{r}^2\rangle_{\text{ac}}$ where $\langle\vec{r}^2\rangle_{\text{ac}}$ is the (active) static variance, Eq.~(\ref{append_MSD-static-active}). The latter diverges with the lower integration limit $\omega_{\text{min}}$, and thus shows a Landau-Peierls-like instability $\langle\vec{r}^2\rangle_{\text{ac}}\sim N^{2/d_s-1}$, similar to the passive case \cite{burioni_2002,burioni_2004}. Hence, for this particular case of force monopoles combined with thermal forces, it is possible to assign an effective temperature 
\begin{equation}
T_{\text{eff}} = T+\frac{\phi {f_0}^2\tau p(1-p)}{d k_B \gamma}, 
\label{teff}
\end{equation}
which can describe both the subdiffusion and saturation regimes of the MSD of an active system. Noteworthy, this applies exactly only for the second moment (i.e., MSD), and likely for higher moments, it is not appropriate since the active noise is Poissonianly distributed (i.e., the random telegraph process) while the thermal noise is Gaussianly distributed \cite{naftali_2022}. An effective temperature is frequently used to characterize active or living systems \cite{eff_temp_loi_2008,eff_temp_loi_2011,eff_temp_review_2011,eff_temp_ghosh_2014}, but clearly, it is not generally applicable (e.g., as shown next for force dipoles).

It is of interest to extend the subdiffusion exponents to the case of a viscoelastic solvent, whose frequency-dependent complex modulus behaves as a power-law, $G^*(\omega)\sim (i\omega)^{\alpha}$, as discussed in Sec. I. This is presented in Appendix \ref{append_sec_viscoelastic}, where we obtain $\nu_{th}=\alpha\left(1-\frac{d_s}{2}\right)$ and $\nu_{ac}=\alpha\left(2-\frac{d_s}{2} \right)-1$, which reduce to the known results for a linear chain, $d_s=1$ \cite{viscoelastic_weber_2010,viscoelastic_vandebroek2014,visco_vandebroek_2015,viscoelastic_polovnikov2018}. Hence, for any fractal in viscoelastic solvent, the two exponents are equal only for $\alpha=1$, the viscous case. Interestingly, for $\alpha<\left(2-\frac{d_s}{2} \right)^{-1}$, the active subdiffusion regime dissapears.

\subsubsection{Force dipoles} For force dipoles -- since for typical fractals $1-d_s/2-d_s/d_{l}<0$ -- {\it anomalous subdiffusion regime is absent}, the MSD crosses over from the early ballistic motion $\sim t^2$ to saturation at $2\langle\vec{r}^2\rangle_{\text{ac}}$. As now the latter diverges (see Eq.~(\ref{append_MSD-static-active})) with the upper integration limit $\omega_0$, it becomes essentially independent of $N$. This implies that, surprisingly, for the combination of thermal and active forces, $\langle\vec{r}^2\rangle_{\text{th}}$ {\it may dominate} the total static MSD for large enough systems due to the Landau-Peierls instability. Moreover, if force monopoles are also present, they are likely to dominate the evolution (as discussed in the next section).
\begin{figure}
\centering
\includegraphics[width=0.435\textwidth,trim={0cm 0.25cm 0.2cm 2.2cm},clip]{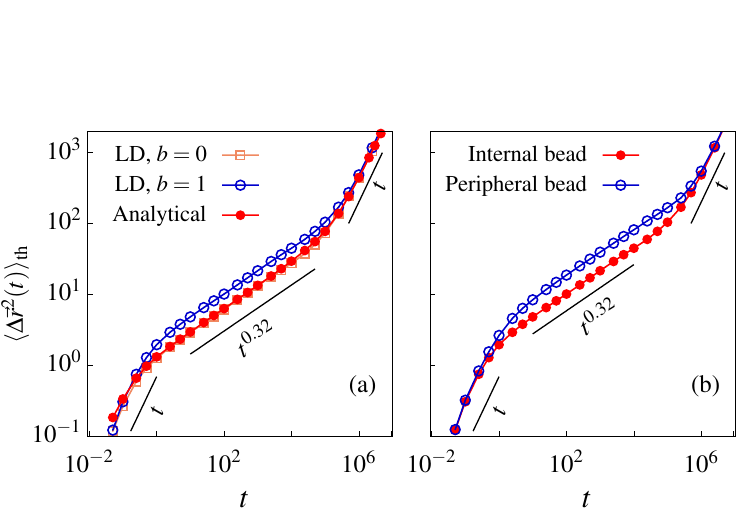}
\caption{{\bf Thermal network.} Mean square displacement (MSD) of a bead in a passive Sierpinski gasket S$_9$ ($N=9843$): (a) MSD of an internal bead, estimated from LD simulations for $b=0$ (square symbol) and $b=1$ (triangle symbol) equilibrium bond length and analytical Eq.~(\ref{MSD_total_th}) (circle symbol) with $d=2$. (b) Comparison of the MSD of two different beads, internal and peripheral, for $b=1$. Curves were fitted to a powerlaw $B_{\text{th}} t^{\nu_{\text{th}}}$ in the time range $5<t<50000$, yielding $\nu_{\text{th}}\simeq 0.32$ in all cases while having different amplitudes (specifically, $\nu_{\text{th}}=0.322 \pm 0.001 $, $0.326\pm 0.001$, $0.327\pm 0.002$ for (a) top to bottom symbols, and $\nu_{\text{th}}=0.326\pm 0.001$, $0.321\pm 0.001$ for (b) top to bottom symbols). }
\label{fig_msd_th}
\end{figure}

\section{Numerical simulations and results}\label{sec_simulation}
These analytical results are based on the scalar elasticity Hamiltonian (Eq.~(\ref{hamiltonian}), assuming a large system, $N\to\infty$. To verify the sensitivity of our results to various assumptions, we performed Langevin dynamics (LD) simulations of a bead-spring Sierpinski gasket (Fig.~\ref{fig_schematic}), using different generations where the $n_{th}$ generation is denoted by S$_n$, with $n=1,2,...$. We solve Eq.~(\ref{Langevin1}) with $H=\frac{1}{2}m\omega_0^2\sum_{<ij>}(\vec{r}_i-\vec{r}_j-b \hat{r}_{ij} )^2$ for the Hamiltonian of harmonic springs having an equilibrium distance $b$, where $\hat{r}_{ij}$ is the unit vector of the distance $\vec{r}_i-\vec{r}_j$ between beads $i$ and $j$. For $b=0$, it reduces to the scalar elasticity Hamiltonian discussed above. We follow the position of an internal bead 
(See Fig.~\ref{fig_gen_9}) in a gasket and calculate its {\it time-averaged} MSD using the standard method for stationary noise. Throughout the simulation, we worked with dimensionless units, where distance is in units of $b$, force in units of $m\omega_0^2 b$, and times are in units of $\tau_0=\gamma/(m\omega_0^2)$, with $k_B T/m\omega_0^2 b^2=1$ (see details in Appendix \ref{append_sec_numerical}). 

\subsection{Thermal} Consider first the dynamics of a thermal (passive) network (previously studied \cite{granek_2005,granek_2011,shlomi_2012_pre,shlomi_2012_prl}). We compare the MSD obtained from simulations, for both vanishing and non-vanishing $b$, with the numerical evaluation of Eq.~(\ref{MSD_total_th}) for S$_9$, $N=9843$. In Fig.~\ref{fig_msd_th}(a), anomalous subdiffusion is observed spanning over three decades with exponents agreeing very well with the predicted theoretical value $\nu_{\text{th}}=1-d_s/2=0.317$ for the Sierpinski gasket with $d_s=1.365$ \cite{alexander_1982,rammal_1983}. 
The MSD amplitude of the analytical model, which is based on the scalar elasticity Hamiltonian, matches well with the LD simulations results for $b=0$, while it differs from the MSD for $b=1$.
The short and long time behavior represents regular diffusion, $r^2\sim t$, such that the short time is the independent bead diffusion while the long represents the CM diffusion.  
For MSD of internal and peripheral beads, we find almost identical exponents but somewhat different amplitudes (see Fig.~\ref{fig_msd_th}(b)), showing the faster motion of the peripheral bead than that of the internal bead.

\begin{figure}
\centering
\includegraphics[width=0.41\textwidth,trim={0cm 0.3cm 0.2cm 2cm},clip]{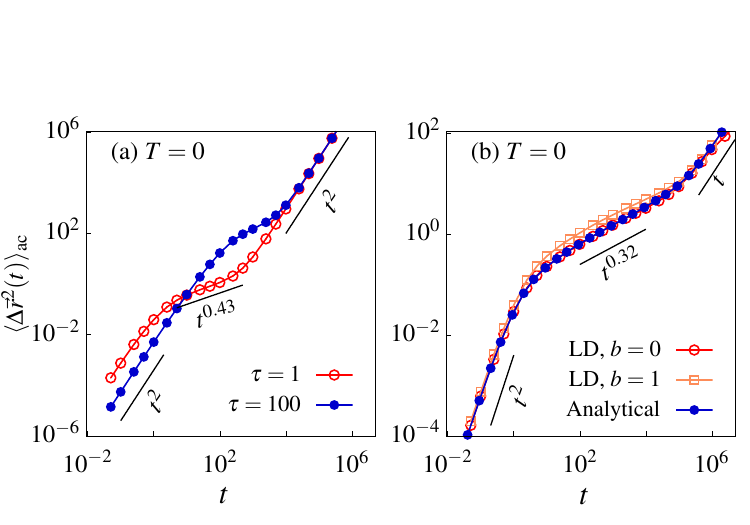}\\
\includegraphics[width=0.41\textwidth,trim={0 0.3cm 0.2cm 2cm},clip]{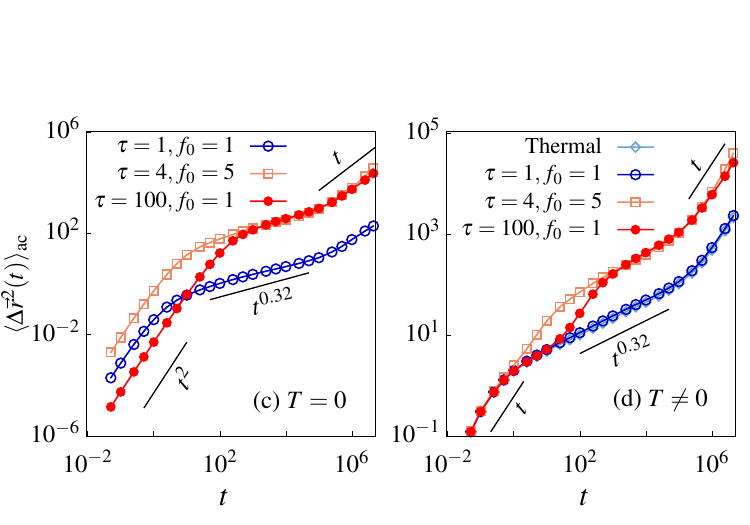}
\caption{{\bf Force monolpoles.} MSD of an internal bead of a Sierpinski gasket S$_9$ ($N=9843$) under stochastic force monopoles at each bead ($\phi=1$) for various active force parameters with `on' probability $p=0.5$, with arrested thermal motion ($T=0$) in (a)-(c). In (a) the MSD is calculated as is, while in (b)-(d) {\it we subtract}, using Eq.~(\ref{append_mono_time_ave_msd}) (see Appendix \ref{append_sec_mono}), {\it the drift motion of the CM}, to suppress the ballistic motion at long times. (a) MSD at different $\tau$ for a given force amplitude $f_0=1$. The long times $t^2$ behavior indicates the ballistic motion of the CM, associated with a drift velocity that emerges due to the finite size of the gasket. (b) MSD estimated from analytical Eqs.~(\ref{MSD_total_ac}) and LD simulations for both vanishing and non-vanishing bond length for $f_0=1$ and $\tau=1$ with the {\it CM drift subtracted}, and the CM diffusion motion is observed at long times. (c) MSD of the active gasket for different force parameters. The subdiffusion exponent $\nu_{\text{ac}}=0.322\pm0.003$ is obtained from a power-law fit in the range [5-50000]. MSD amplitudes, Eqs.~(\ref{superdiff})-(\ref{subdiff}): $B_{acs}=0.0678, 0.7132, 0.0034$ and $B_{ac}=0.247, 17.3, 20.3$, for $(\tau, f_0)=(1,1), (4,5), (100,1)$, respectively.
(d) MSD of the active gasket including thermal fluctuation, $T\neq 0$. The combination of active and thermal (passive) contributions to the MSD is compared with the purely thermal MSD. $r$ is in unit of $b$, $f_0$ in unit of $m\omega_0^2 b$, and $t$ and $\tau$ are in units of $\tau_0=\gamma/(m\omega_0^2)$.}
\label{fig_mono_msd}
\end{figure}

\subsection{Force monopoles} Turning to active fractal networks with force monopoles, to each bead we assign an active force of random orientation. In Fig.~\ref{fig_mono_msd}(a) and (c) (drift velocity eliminated, see below), we depict the MSD of an internal bead for different values of active force parameters and arrested thermal motion ($T=0$). We observe a crossover at $t\sim \tau$, from short-time ballistic motion, MSD $\sim t^2$, to subdiffusion at intermediate times. In  Fig.~\ref{fig_mono_msd}(a), a crossover to ballistic motion appears at longer times, indicating the presence of a CM drift velocity. The latter emerges from the incomplete cancellation of the total force due to the system's finite size. For $N\to\infty$, $\sum_{i\neq j}\hat f_i\cdot\hat f_j=0$; however, in a finite system, this sum may have a residual (random) value, resulting in drift velocity $v_{\text{drift}}\sim \frac{\phi f_0 p}{\gamma \sqrt{N}}$ (Appendix \ref{append_sec_mono}) which will nevertheless vanish if we average over many realizations of the force field.

\subsubsection{Subdiffusion} Calculating the MSD about the drift position using Eq.~(\ref{append_mono_time_ave_msd}) (Appendix \ref{append_sec_mono}), we show in Fig.~\ref{fig_mono_msd}(c) that the intermediate-time subdiffusion regime becomes longer, MSD$\sim t^{\nu_{\text{ac}}}$ with $\nu_{\text{ac}}\simeq 1-d_s/2 \simeq 0.32$ as analytically predicted, and for $t \gg \tau_N$ the long-time ballistic motion is replaced with $\sim t$ behavior associated with CM active diffusion. We can also observe the effects of the correlation time $\tau$ and force amplitude $f_0$ on the MSD ballistic and subdiffusion amplitudes confirming our prediction $B_{acs}\propto \phi f_0^2\tau^{-0.682}$ ($t\ll \tau$) and $B_{ac}\propto \phi f_0^2\tau$ ($t\gg \tau$). Like the passive case, for $b=0$, theory and simulations match exactly, see Fig.~\ref{fig_mono_msd}(b); for $b=1$, the subdiffusion amplitude somewhat differs, yet the exponent is unaltered.

\subsubsection{Combined active and thermal motion} 
In Fig.~\ref{fig_mono_msd}(d), we show the combined 
effect of active and thermal forces for a few sets of parameters and compare them with the thermal MSD. When thermal forces are present, at short times $t<\tau_0$, the thermal contribution dominates resulting in diffusive behavior. At intermediate times, subdiffusion regimes with $\nu\simeq 1-d_s/2 \simeq 0.32$ can be observed for all cases; in one combination, having a long $\tau$ ($=100 \tau_0$) and small $f_0$ ($=m\omega_0^2 b$), two consecutive such regimes appear, where the first is dominated by thermal motion and is followed by a second one, dominated by activity and having a much larger amplitude.
\begin{figure}
\centering
  \includegraphics[width=0.3\textwidth,trim={1.3cm 0.6cm 1.2cm 0.7cm},clip]{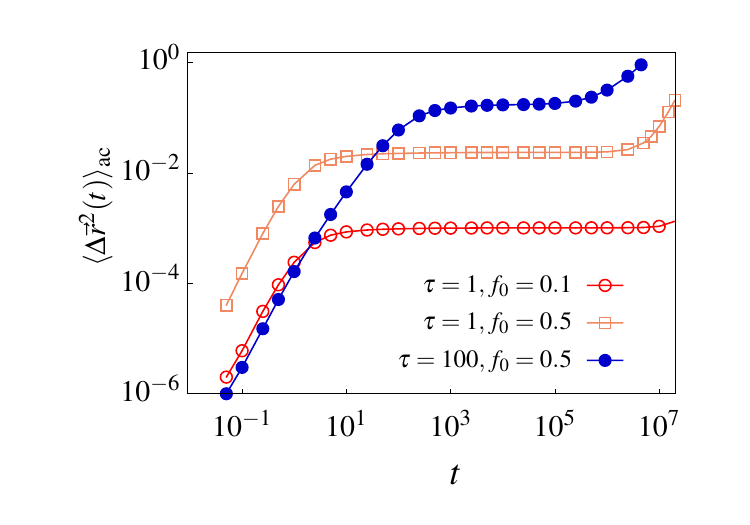}
\caption{{\bf Force dipoles.} MSD of an internal bead of gasket S$_9$ ($N=9843$) subjected to stochastic force dipoles in the absence of thermal motion. The force dipoles are randomly assigned to $\phi=20\%$ of the network bonds. $r$ is in unit of $b$, $f_0$ in unit of $m\omega_0^2 b$, and $t$ and $\tau$ are in units of $\tau_0=\gamma/(m\omega_0^2)$.}
\label{fig_dipole_gen_9}
\end{figure}

\subsection{Force dipoles} Next, we simulate force dipoles randomly assigned to a fraction $\phi=20\%$ of total bonds (unless otherwise specified), where we either prohibit or do not prohibit dipoles having a common bead. In biological networks, the choice of distinct beads may arise due to the excluded volume interaction between motor proteins. During the simulations, we update each dipole orientation to remain parallel to $\vec{r}_{ij}$. 

\subsubsection{Absence of subdiffusion} 
We estimate the MSD for gasket S$_9$ (Fig.~\ref{fig_dipole_gen_9}) for different force parameters. For times $t \ll \tau$, the MSD shows a ballistic behavior, $\sim t^2$, which saturates to a constant value at $t \sim \tau$. Both the existence of the early ballistic motion and the {\it absence} of the intermediate subdiffusion regime for the Sirepinski gasket where $1-d_s/2-d_s/d_l=-0.5464<0$, are in accord with the analytical predictions.

\subsubsection{Long-time rotational motion} In Fig.~\ref{fig_dipole_gen_9}, at longer times, a ballistic-like rise in MSD for higher $f_0^2\tau$ is observed, which requires further analysis.
However, at longer times, a ballistic-like rise in MSD for higher $f_0$ is observed in Fig.~\ref{fig_dipole_gen_9}, followed by {\it oscillations} that can be picked for smaller gaskets, see Fig.~\ref{fig_dipole_multi}(a) for S$_4$. In the movies (snapshots displayed in Fig.~\ref{sm1}-\ref{sm3} in Appendix, Multimedia available online) showing trajectories of a few gasket generations, we can observe a {\it persistent rotational motion} of the network, either clockwise or anti-clockwise, that may explain the MSD oscillatory behavior. To confirm this, we define a mean square angular displacement (MSAD) associated with the pure rotational motion of an arbitrary bead $i$, located at mean square distance $\langle R^2\rangle$ from the CM, $\langle R^2\rangle\langle(\hat{n}(t)-\hat{n}(0))^2 \rangle$, where $\hat{n}(t)$ is the unit vector of $\vec{R}$. In Fig.~\ref{fig_dipole_multi}(a), we plot the MSAD together with the complete MSD; the inset (linear scale) emphasizes the motion at long times. The overlap of the two oscillatory curves is almost perfect. Importantly, even though force dipoles cannot generate a net torque, they can induce a mean rotational velocity (see discussion in Appendix \ref{append_sec_dipole}). 

\begin{figure}
\centering
\hspace{-0.1cm}\includegraphics[width=0.37\textwidth,trim={0.1cm -0.2cm 2cm 0.01cm},clip]{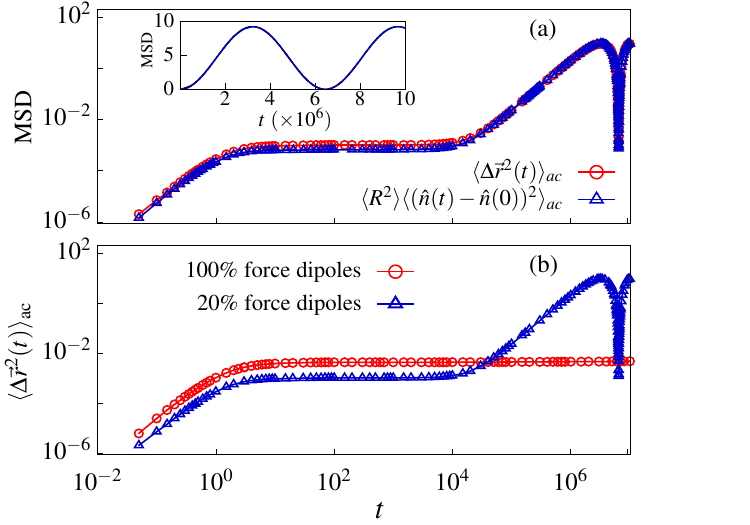}
\caption{{\bf Force dipoles.} MSD of an internal bead of gasket S$_4$ ($N=42$) subjected to stochastic force dipoles in the absence of thermal motion. (a) MSD $\langle \Delta r^2(t)\rangle$, and mean square angular displacement (MSAD), $\langle R^2\rangle \langle(\hat{n}(t)-\hat{n}(0))^2\rangle$, of an internal bead with force dipoles randomly assigned to $\phi=20\%$ of the network bonds. The inset shows both curves on a linear scale. (b) MSD of an internal bead of S$_4$ for $\phi=20\%$ bonds (blue color) and $\phi=100\%$ bonds (red color) associated with force dipoles. The force parameters are: $\tau=1.0$, $p=0.5$, and $f_0=0.1$. $r$ is in unit of $b$, $f_0$ in unit of $m\omega_0^2 b$, and $t$ and $\tau$ are in units of $\tau_0=\gamma/(m\omega_0^2)$.}
\label{fig_dipole_multi}
\end{figure}

\begin{figure*}
\centering
\hspace{1cm}\includegraphics[width=4cm]{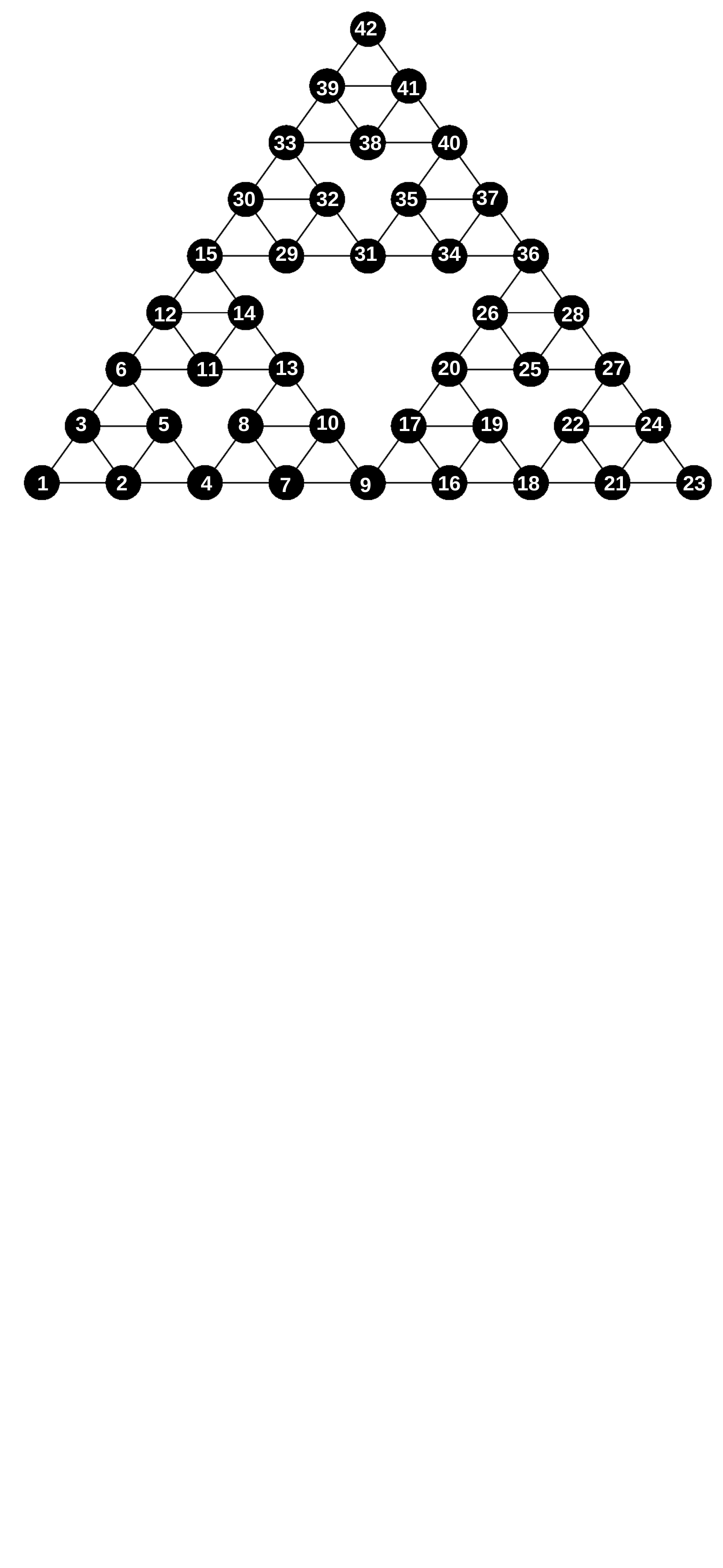}
\hspace{-5cm}\includegraphics[width=0.63\textwidth]{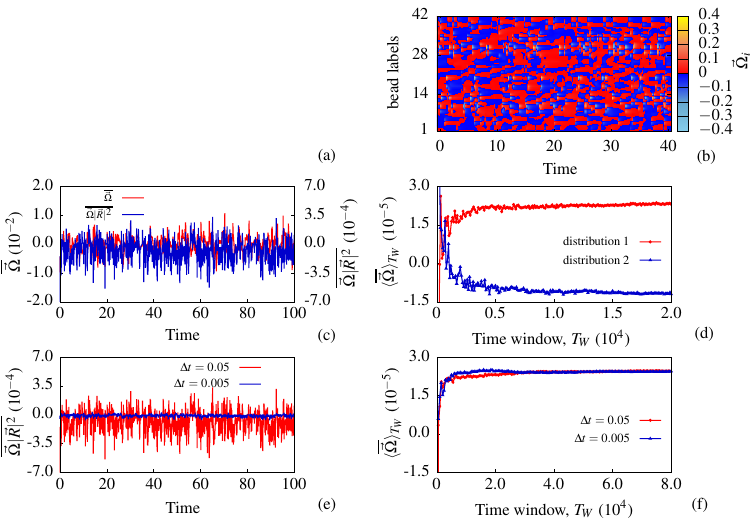}
\caption{{\bf Force dipoles.} Rotational motion of Sierpinski gasket S$_4$ subjected to stochastic force dipole with arrested thermal motion ($T=0$). (a) Schematic of S$_4$ ($N=42$) with indices from $1$ to $42$ assigned to different beads. (b) Color map showing the time-dependent angular velocity $\vec{\Omega}_i$ fluctuations of each bead. (c) Temporal Mean angular velocity $\overline{\vec{\Omega}}$ (Eq.~(\ref{mean_ang_vel})) and the mean radius-square-weighted angular velocity $\overline{\vec{\Omega} |\vec{R}|^2}$ (Eq.~(\ref{mean_ang_vel_r2})), averaged over beads at different times. (d) The time average of mean angular velocity, $\langle\overline{\vec{\Omega}}\rangle_{T_W}$ for different time-averaging windows, $T_W$, converging to a non-vanishing constant value for long values of TW
that depends on the distribution of the dipolar forces. (e) $\overline{\vec{\Omega}|\vec{R}|^2}$ for different simulation time step $\Delta t$. The fluctuations seen in $\overline{\vec{\Omega}|\vec{R}|^2}$ improve with a smaller $\Delta t$. 
(f) $\langle\overline{\vec{\Omega}}\rangle_{T_W}$ for different simulation time step $\Delta t$. The difference between the two curves is minor and effectively disappears when the time-averaging window gets long. The force parameters are: $\tau=1.0$, $p=0.5$, and $f_0=0.5$. $r$ is in unit of $b$, $f_0$ in unit of $m\omega_0^2 b$, and $t$ and $\tau$ are in units of $\tau_0=\gamma/(m\omega_0^2)$.
}
\label{fig_ave_ang_vel}
\end{figure*}

To investigate the ``persistent" rotational motion of the entire network, we compute the angular velocity $\vec{\Omega}_i$ of individual beads and subsequently determine the mean angular velocity (averaged over beads) $\overline{\vec{\Omega}}$ (Eq.~(\ref{mean_ang_vel})). The color map (Fig.~\ref{fig_ave_ang_vel}(b)) displays the temporal fluctuation of angular velocity $\vec{\Omega}_i$ for individual beads. Similarly, fluctuation of $\overline{\vec{\Omega}}$ over time is shown in Fig.~\ref{fig_ave_ang_vel}(c). However, the time average of mean angular velocity, $\langle\overline{\vec{\Omega}}\rangle_{T_W}$ for different time averaging windows, $T_W$ (see Fig.~\ref{fig_ave_ang_vel}(d)) is found to be non-vanishing and converges to a constant value for long values of $T_W$ which depends on the (quenched) realization of the dipolar forces. Given the local fluctuation in $\overline{\vec{\Omega}}$, the non-zero value of $\langle\overline{\vec{\Omega}}\rangle_{T_W}$ results in a ``crawling" (non-coherent) rotation of whole object which appears as persistent rotational motion at long time windows. 

Condition of net zero torque requires that mean radius-square-weighted angular velocity is zero, i.e., 
$\overline{\vec{\Omega} |\vec{R}|^2}=0$ (Eq.~(\ref{mean_ang_vel_r2})) which does {\it not} necessarily imply, $\overline{\vec{\Omega}} = 0$. 
Here, $\vec{R}_i$ is the position of $i$-th bead in the CM frame of reference.
The time-dependent fluctuations of $\overline{\vec{\Omega} |\vec{R}|^2}$ presented in Fig.~\ref{fig_ave_ang_vel}(c), is found to be {\it two orders of magnitude smaller} than that of 
$\overline{\vec{\Omega}}$, despite that for all beads $|\vec{R}_i|^2>1$.  The fluctuations seen in 
$\overline{\vec{\Omega}|\vec{R}|^2}$ reflect the minor simulation errors, which improve with a smaller simulation time step, $\Delta t$. When the simulation time step is 10-fold decreased, these fluctuations get significantly smaller, shown in Fig.~\ref{fig_ave_ang_vel}(e). Nevertheless, in Fig.~\ref{fig_ave_ang_vel}(f), we show that the difference in $\langle\overline{\vec{\Omega}}\rangle_{T_W}$ between the two simulations (of different time steps) is minor and effectively vanishes when the time averaging window gets long, demonstrating that the ``crawling"  rotation of the whole object is real.

We further argue that this rotational motion, whose direction and frequency randomly vary between realizations of the dipole spatial distribution (Fig.~\ref{fig_dipole_multi_si}(a) in Appendix \ref{append_sec_dipole}), arises from a small residual anisotropy caused by the asymmetric distribution of forces associated with finite-size effects. 
As shown in Fig.~\ref{fig_dipole_multi_si}(c), the onset of oscillations is delayed to 
longer times as $N$ increases, reflecting a reduction in anisotropy. Furthermore, 
simulating a gasket where force dipoles are assigned to all bonds, completely eliminates the long-term MSD oscillations, as shown in Fig.~\ref{fig_dipole_multi}(b).

Interestingly, in {\it triangular micro-swimmers} that are subject to force dipoles and hydrodynamic interaction, rotational motion emerges too \cite{swimmer_rizvi2018,rizvi_2020}. 
In Fig.~\ref{fig_triangle} (see Appendix \ref{append_sec_numerical}), we show that even Rouse dynamics 
(i.e., without hydrodynamic interaction) of a triangular bead-spring, with anisotropy of force dipole strength (and random-telegraph fluctuations), is enough to produce rotational motion. 

In fact, it has been suggested in general that, despite the vanishing total torque, a ``crawling" rotational motion -- associated with a given sequence of deformations -- could be expected for a {\it deformable} body, once the rotational symmetry of (internal) forces, i.e., the force dipoles, is broken \cite{rotational_shapere_1987,rotational_shapere_1989,rotational_shapere_1989geometry,rotational_raz_2008}, similar to translational swimming in low Raynold numbers.
The broken symmetry due to force dipole distribution and the system’s geometry can create localized deformations that can spread through the system in a coordinated manner. This sequence of deformations (shape changes) can cause the body to ``crawl” or exhibit a slow, persistent, and directional (albeit non-torque-driven) rotational motion.
\begin{figure}
\centering
\includegraphics[width=0.375\textwidth,trim={0.03cm 0 0.2cm 0.1cm},clip]{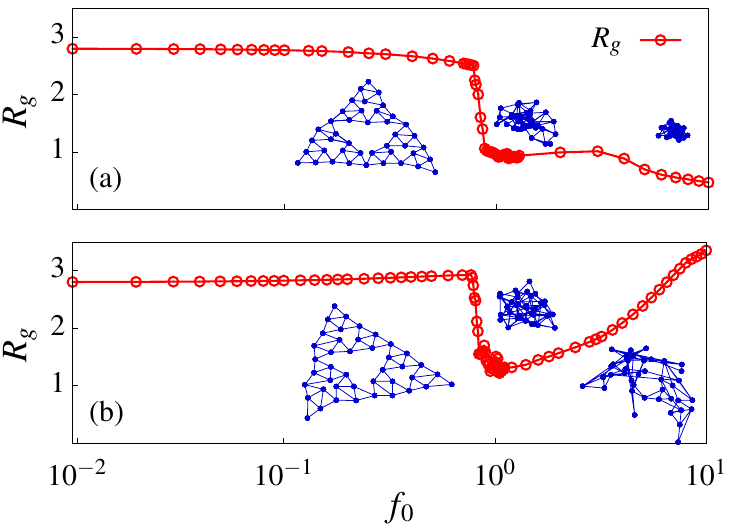}
\caption{{\bf Force dipoles:} Radius of gyration $R_g$ of Sierpinski gasket S$_4$ (on semi-log plot) at different $f_0$ for 
$\tau=\tau_0$ and $p=0.5$; (a) contractile, and (b) extensile force dipoles. Snapshots of the gasket at different $f_0$ are shown for both cases. $R_g$ and $f_0$ are in units of $b$ and $m\omega_0^2 b$, respectively.
}
\label{fig_rg}
\end{figure}

\subsubsection{Network collapse} 
\begin{figure}
\centering
\includegraphics[width=0.385\textwidth,trim={0 0.06cm 0.0cm 2.6cm},clip]{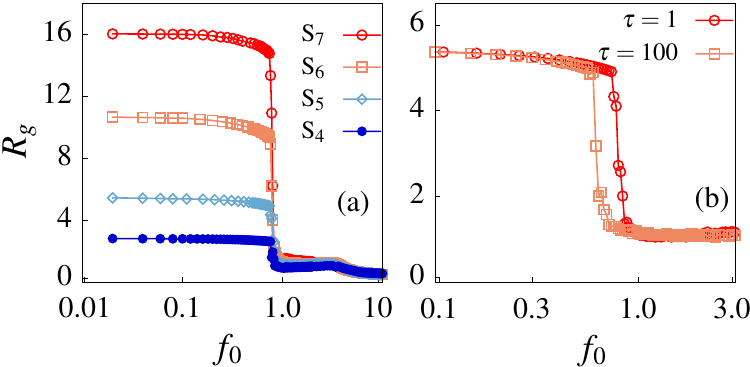}
\caption{{\bf Force dipoles:} Radius of gyration $R_g$ of Sierpinski gasket (on semi-log plot) 
at different $f_0$ for contractile force dipoles with $p=0.5$; (a) for different generations, at given $\tau=\tau_0$, and (b) at different $\tau$ for gasket S$_5$. $N=42, 123, 366, 1095$ for S$_4$, S$_5$, S$_6$ and S$_7$, respectively. $R_g$, $f_0$ and $\tau$ are in units of $b$, $m\omega_0^2 b$ and $\tau_0$ respectively.}
\label{fig_rg_tau}
\end{figure}

The above behavior for force dipoles does not apply above a {\it critical force} amplitude $f_{0c}\approx m\omega_0^2 b$. For $f_0>f_{0c}$, the gasket {\it collapses} to a random shape maintaining dynamical fluctuations, but with rotational motion effectively arrested (see Fig.~\ref{fig_dipole_MSAD_force}). The radius of gyration $R_g$ for contractile force dipoles against $f_0$ is shown in Fig.~\ref{fig_rg}(a) for S$_4$, and in Fig.~\ref{fig_rg_tau}(a) for generation S$_5$$-$S$_7$. The transition persists and sharpens for larger (higher generation) gaskets, implying a true, activity-induced, first-order phase transition.

The collapse may be rationalized as a result of the dynamical ``persistence length" \cite{dynamical_persistence_RMP_2016,dynamical_persistence_cao_2019} -- generalized to account for the spring resistance -- rising above $b$, leading to 
$f_{0c}\simeq m\omega_0^2 b/\left(1-e^{-\tau/\tau_0}\right)$ (see Appendix \ref{append_sec_persistence}), in agreement with our results for $\tau\gtrsim \tau_0$, Fig.~\ref{fig_rg_tau}(b). Contrary to contractile force dipoles 
(Fig.~\ref{fig_rg}(a)), for extensile force dipoles $R_g$ rises again at larger forces, Fig.~\ref{fig_rg}(b).

\section{Application to Chromatin Dynamics}\label{sec_chromatin}
To connect with chromatin dynamics studies described in Ref.\cite{weber_2012}, we
need to establish the parameter regime associated with chromatin. We hypothesize that the
chemical distance between crosslinks in chromatin is equal to the persistence length
$L_p\simeq 250$ nm \cite{lp_250_gero_2002,lp_250_kerstin_2004}, and that $\tau$ and $f_o$ are associated with cytoplasmic motor protein processivity times and forces. Cytoplasmic typical values are in the range $f_0=1-10 pN$, $\tau=10 ^{-3} - 10s$
\cite{parameter_cui_2000,parameter_JOHNSTONE_2020,parameter_fierz_2019,parameters_zhou_2016}.
To obtain the dimensionless parameter range applicable to chromatin dynamics, we assume the
network segment between two beads behaves as a Gaussian 2D thermal spring with spring constant
$m\omega_0^2=\frac{90k_B T L_p^2}{b^4}$ for semi-flexible polymers \cite{mackintosh_1995elasticity,mackintosh_2014_RPM}, and the bond length between two beads
is $b=L_p$. Taking the solvent viscosity (water) $\eta=1\times10^{-3}$Pa.s, we estimate $m\omega_0^2b = \frac{90k_BT}{b} = 1.5 \, pN$ and $\tau_0= \frac{\gamma}{m\omega_0^2} = \frac{3\pi\eta b^3}{90k_BT} = 4 \times 10^{-4}\, s$. Hence, for chromatin, the dimensionless values of force amplitude and force
correlation time are in the range $f_0/(m\omega_0^2 b)=0.5-7$ and $\tau/\tau_0=1-10^4$, respectively. These parameters yield MSDs of passive systems and purely active (i.e., $T=0$) force monopole systems, exhibiting subdiffusion regime with {\it identical exponents} yet different amplitudes, as shown in Fig.~\ref{fig_mono_msd}(b)-(c).

To describe realistically untreated cells, we should consider the interplay of thermal forces (as white noise) with both active monopolar and dipolar forces.
Moreover, with $\nu_{\text{th}}=\nu_{\text{ac}}=1-d_s/2\simeq 0.4$ (the value observed in Ref.~\cite{weber_2012}), we can predict $d_s\simeq 1.2$, associated with a slightly cross-linked chain. This deviates from the original fractal globule model \cite{grosberg_1988} that ignores the presence of such crosslinks and corresponds to the case where $d_s=1$. Indeed, Rouse dynamics of passive and force-monopole-active linear Gaussian chains do yield $\nu_{th}=\nu_{ac}= 1/2$, consistent with our result with $d_s=1$ \cite{osmanovic_2017,winkler_2020}. Our deduction of $d_s\simeq 1.2$ does agree with experimental findings showing slight deviations from the fractal globule model predictions for the static correlations and fractal dimension \cite{fractal_bancaud_2009}. Furthermore, Bronshtein {\it et al.} demonstrated the role as crosslinkers of {\it lamin A} proteins in Eucrayotic cells chromatin \cite{lamina_a_garini_2015}.

\begin{figure}
\centering
\includegraphics[width=0.47\textwidth,trim={0cm 0.4cm 0cm 0.5cm},clip]{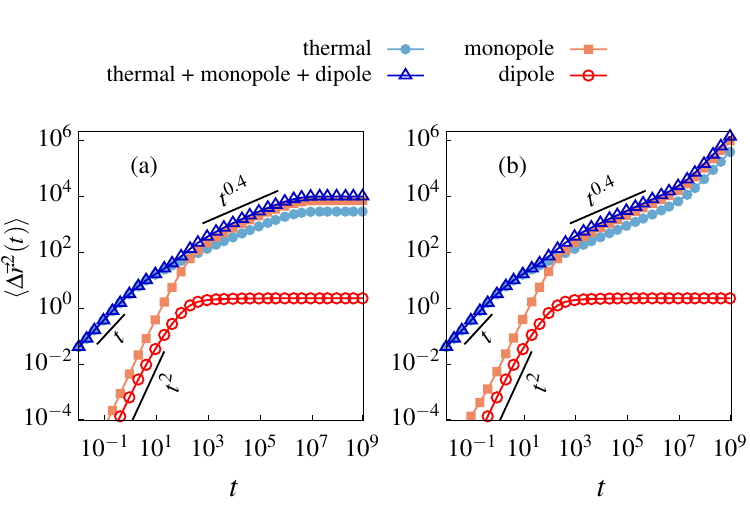}
\caption{{\bf Chromatin-like active network with force monopoles and dipoles both included:} 
MSD of an arbitrary bead of a fractal network ($N=9843$) with spectral dimension $d_s=1.2$, subject to active forces (both dipolar and monopolar forces) and thermal forces, compared to motion with different force sources individually; 
(a) MSD without the CM motion, and (b) MSD with CM motion. 
The MSD is estimated from the numerical integration of Eq.~(\ref{append_two_point_integral}) with $\omega_{\text{min}}=\omega_0 N^{-1/d_s}$ for the lower limit, using the spectral dimension $d_s=1.2$, that is predicted from our theoretical analysis $\nu=1-d_s/2$ considering the experimental value of subdiffusion exponent $\nu=0.4$ for chromatin in ref\cite{weber_2012}. For active forces, the force parameters are $f_0=1$, $\tau=100$, $p=0.5$, 
and $\phi=0.2$ for both monopolar and dipolar forces, such that the relative strength of active forces (variance) to thermal forces (variance) for the network is 
$\phi f_0^2 \tau p(1-p)/d=2.5$ (in dimensionless unit). $r$ is in unit of $b$, $f_0$ in unit of $m\omega_0^2 b$, and $t$ and $\tau$ are in units of $\tau_0=\gamma/(m\omega_0^2)$.} 
\label{fig_mixed_active_integration}
\end{figure}

\begin{figure}
\centering
\includegraphics[width=0.35\textwidth,trim={0.6cm 0.2cm 0.9cm 0.7cm},clip]{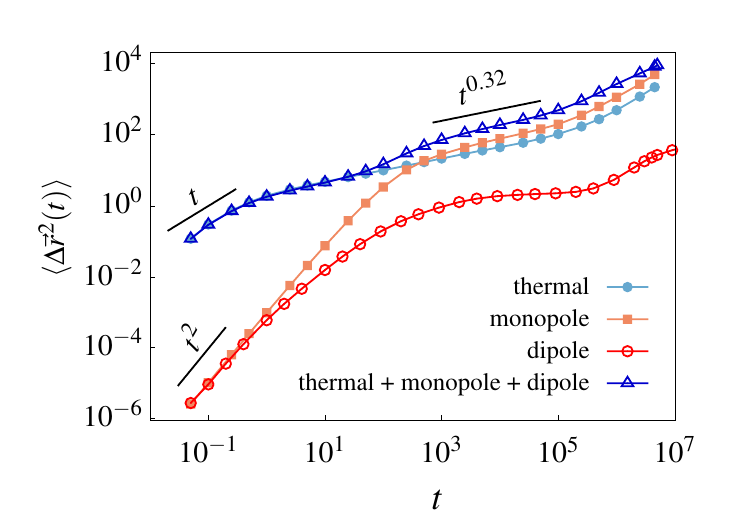}
\caption{{\bf Active Sierpinski gasket with chromatin-like force monopoles and dipoles from LD simulations:} MSD of an internal bead of Sierpinski gasket S$_9$ ($N=9843$) subjected to active forces (both dipolar and monopolar) and thermal forces, compared to motion with different force sources individually. Here, 
the MSD is estimated from LD simulations, and the force parameters are the same as in Fig.~\ref{fig_mixed_active_integration}.} 
\label{fig_mixed_active}
\end{figure}

In Fig.~\ref{fig_mixed_active_integration}, we present the combination of thermal, monopolar, and dipolar forces (with comparison to the individual force source cases) for a fractal network ($N\simeq 10^4$) 
with spectral dimension $d_s=1.2$. In Fig.~\ref{fig_mixed_active_integration}(a), we estimate the MSD from the numerical integration of Eq.~(\ref{append_two_point_integral}), which does not include the CM motion, using $\omega_{\text{min}}=\omega_0N^{-1/d_s}$ for the lower limit and the spectral dimension $d_s=1.2$ as we suggest for chromatin. In Fig.~\ref{fig_mixed_active_integration}(b) we add the CM motion to Eq.~(\ref{append_two_point_integral}) according to Eqs.~(\ref{th_CM_motion})-(\ref{ac_CM_motion}). We can still observe a significant subdiffusion regime described by an exponent $\nu\simeq 1-d_s/2=0.4$, which emerges since the dipolar sources alone lead only to a moderate MSD saturation value (denoted as $2\langle\vec{r}^2\rangle_{\text{ac}}$ in Eq.~(\ref{append_MSD-static-active-1})) allowing the combination of thermal and monopolar forces to take over. For comparison, in Fig.~\ref{fig_mixed_active}, we show the equivalent MSD results for a Sierpinski gasket ($N=9843$) from LD simulations. While the early evolution is similar to Fig. 8 (except for a different subdiffusive exponent), for force dipoles alone we also observe the rise from the plateau at long times (associated with the slow rotation), which is not captured by the analytic theory. However, it is clear that the dipolar forces do not contribute to the subdiffusive motion induced by the active monopolar and thermal forces. Hence, we can assess that monopolar and thermal forces are the cause of the subdiffusive motion in chromatin.

\begin{figure}
\centering
\includegraphics[width=0.47\textwidth,trim={0cm 0.4cm 0cm 0.5cm},clip]{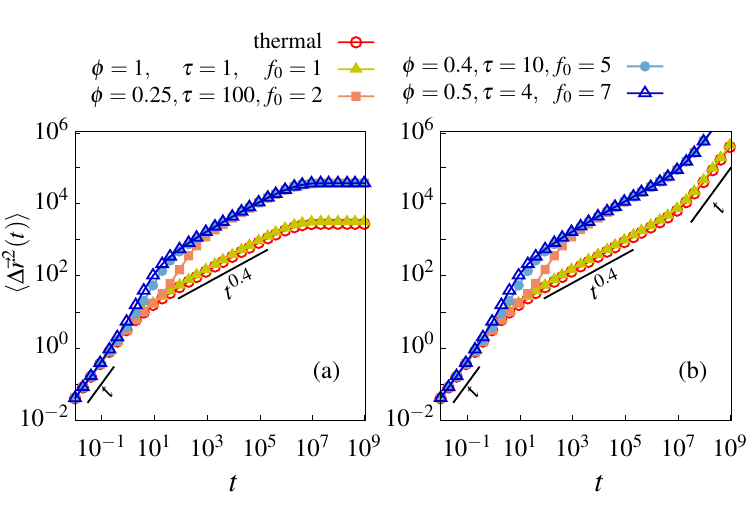}
\caption{{\bf Chromatin-like active network with force monopoles:} 
MSD of an arbitrary bead of a fractal network ($N=9843$) with spectral dimension $d_s=1.2$, subjected to active monopolar forces and thermal forces together, compared to motion with thermal force alone for different force parameters; (a) MSD without CM motion, and (b) MSD with CM motion. 
The MSD is estimated from the numerical integration of Eqs.~(\ref{append_two_point_integral}), same as in Fig.~\ref{fig_mixed_active_integration}. 
We note that the relative strength of active forces (variance) to thermal forces (variance) for the network is $\phi f_0^2 \tau p(1-p)/d$ (in dimensionless unit). $r$ is in unit of $b$, $f_0$ in unit of $m\omega_0^2 b$, and $t$ and $\tau$ are in units of $\tau_0=\gamma/(m\omega_0^2)$.}      
\label{fig_chromatin}
\end{figure}

To examine further the predicted MSD amplitudes (or 'apparent diffusion coefficients') in comparison with the experimental results \cite{weber_2012}, we consider in Fig.~\ref{fig_chromatin} the combination of thermal forces and active force monopoles in comparison to the purely thermal case (with dipolar forces ignored), for a chromatin-like fractal network 
where $d_s=1.2$. 
For small values of $\phi f_0^2 \tau$ where $\phi f_0^2\tau p(1-p)/(d\gamma k_B)\ll T$, i.e., $T_{\text{eff}}\simeq T$, the thermal contribution to the MSD dominates over the active contribution. For higher values of $\phi f_0^2 \tau$ where $T_{\text{eff}}\gtrsim T$, we observe the appearance of either a single subdiffusion regime dominated by active motion 
(for relatively small $\tau$), or a thermally dominated subdiffusive regime crossing over to an actively dominated subdiffusive regime with the same exponent $\nu$, albeit with a higher amplitude $B_{\text{eff}}$. Importantly, this is quite similar to the observation in chromatin reported in Ref. \cite{weber_2012} (see, e.g., Fig. 1 therein). The experimentally measured MSD amplitude ratios, $B_{\text{th}}/B_{\text{eff}}$, of ATP-depleted to normal cells \cite{weber_2012} (Table 1 therein) are theoretically reproduced, depending on the combination $\phi f_0^2\tau$. For example, for {\it S. cerevisiae}, we need $\phi f_0^2\tau\simeq 100$ to match the theoretical ratio with the experimental one, $B_{\text{th}}/B_{\text{eff}}\simeq 0.12$, and for {\it E. coli} we require $\phi f_0^2\tau\simeq 13$ to match the measurements on {\it E. coli}.


Some other recent MSD measurements of chromosomal locus in normal Eukaryotic-cells 
(i.e., no ATP depletion), either single-point MSD or two-point MSD \cite{paper_suggested_stochastic_2023,paper_suggested_folding_2022,paper_suggested_looping_2022}, 
do show subdiffusion exponent, $\nu$ values closer to $1/2$, and fractal globule packing with $d_f=3$ \cite{paper_suggested_stochastic_2023}. Importantly, these results are in perfect accord with our general approach. Provided that $d_s=1$, as is the case for fractal globule, the anomalous diffusion exponent $\nu=1-d_s/2$ will always be 1/2, regardless of the fractal dimension $d_f$. It remains unknown whether ATP-depleted cells in such systems also exhibit the same exponent, a significant observation as we discussed. In case they would, for example, our general approach would enable us to infer that $d_s=1$ in these cells, as in the original fractal globule model. To draw such a conclusion, it is imperative to have MSDs of both ATP-depleted and normal cells, where the values of $\nu$ are identical. This highlights how MSD measurements provide insights into the abundance of crosslinks in chromatin.

Recent active, dipolar-force-based models for chromatin dynamics also appear to yield identical exponents, yet with $\nu_{th}\simeq\nu_{ac}\simeq 0.5$ \cite{dipole_Brahmachari_2023,dipole_put_2019}. In Ref. \cite{dipole_Brahmachari_2023}, the active system is studied at finite temperatures, and the subdiffusion is found only for weak forces, such that the system is thermally dominated \footnote{see Fig. 2F in Ref. \cite{dipole_Brahmachari_2023}, light-blue curve, and long times}. When the system becomes dominated by stronger forces \footnote{see Fig. 2F in Ref. \cite{dipole_Brahmachari_2023}, other colors}, the MSD shows a crossover from super-diffusion at times shorter than the correlation time, to a constant value at time longer than the correlation time. Notably, a similar behavior is obtained in the present work only for dipolar forces.

In Ref. \cite{dipole_put_2019}, where the time window where anomalous subdiffusion was found numerically in the pure active case, $0.05<t <0.30$ \footnote{Fig. 3, left panel, in Ref. \cite{dipole_put_2019}}, is less than one decade and is likely picking the crossover behavior from the $\sim t^2$ to a constant value. This crossover from $t^2$ to a constant value is qualitatively identical to our findings for dipolar forces (e.g., Fig.~\ref{fig_dipole_multi}). In contrast, our sub-diffusion regime for force-monopoles ranges over a few decades (Fig.~\ref{fig_mono_msd}), demonstrating a true anomalous regime. Ref. \cite{dipole_put_2019} claims to derive analytically $\nu=1/2$ for force dipoles (to support the numerical study) and $\nu=3/2$ for force monopoles, both for times shorter than the force correlation time. In contrast, our study yields $\nu=2$ for both force monopoles and dipoles in this time range.

To conclude this section, we comment again on the inability of the ``power-law fluid" interpretation, $G^*(\omega)\sim (i\omega)^{\alpha}$, to explain the observed equality of the active and thermal exponents on the same footing\cite{viscoelastic_weber_2010,weber_2012}. Our extended exponents $\nu_{th}=\alpha\left(1-\frac{d_s}{2}\right)$ and $\nu_{ac}=\alpha\left(2-\frac{d_s}{2} \right)-1$, imply that $\nu_{ac}\neq \nu_{th}$ for any $d_s$, unless $\alpha=1$ (viscous fluid). Taking\cite{viscoelastic_weber_2010} $d_s=1$ and $\alpha$ in the range 0.7-0.8, implies $\nu_{th}=0.35-0.4$ and $\nu_{ac}=0.05-0.2$, which could be well discriminated experimentally. Furthermore, for $d_s=1$ and $\alpha<2/3$, one obtains $\nu_{ac}=0$, i.e., the active subdiffusion is absent, and the MSD saturates to a constant value or crosses over to the CM diffusion.

\section{Discussion and Conclusions}\label{sec_discussion}
We studied analytically arbitrary fractal networks that obey harmonic scalar elasticity and Rouse-type dynamics and are subject to active force-monopoles and force-dipoles. We also performed Langevin dynamics simulations of the Sierpinski gasket considering a nonzero equilibrium spring length for which the elasticity is not perfectly scalar. 
Our results show {\it the absence of a true power-law subdiffusion regime} when only force dipoles are at play within an ideal fractal network. However, when thermal forces, active force monopoles, or both are introduced into the system, their impact on MSD surpasses that of the dipoles, with the latter contributing only a constant saturation value at intermediate times. The resulting subdiffusion exponent is always $\nu=1-d_s/2$ irrespective of whether thermal forces, active force-monopoles, or a combination of 
both predominate the dynamics. The LD simulations confirm the theory with high accuracy and show that the non-vanishing spring length does not alter this exponent. 
The value of the exponent $\nu$ depends solely on the spectral dimension $d_s$ and is independent of the fractal dimension $d_f$ of the self-similar structure. 
In particular, for a linear chain that folds in space without forming internal crosslinks, we always have $d_s=1$ and $\nu=1/2$, regardless of whether $d_f=2$ for Gaussian chains, $d_f=5/3$ for self-avoiding chains, or $d_f=3$ for compact folding.

Thereby, we explain here the apriori surprising result of identical subdiffusion exponent for both normal and ATP-depleted cells. Moreover, we showed that within the range of force strengths ($f_0$), force correlation times ($\tau$), and force density ($\phi$), yielding different combinations of $\phi f_0^2\tau$, we obtain amplitude ratios, $B_{\text{th}}/B_{\text{eff}}$, of ATP-depleted to normal cells that are similar to those experimentally observed \cite{weber_2012}. We conclude that the observed normal cell chromosomal-loci subdiffusion is caused by the combination of active force monopoles and thermal forces. However, we do not claim that dipole forces are absent; they simply do not contribute to this observable. We stress that our conclusions do not rely on the type of self-similarity for chromatin packing, in particular, whether it obeys the fractal globule model \cite{grosberg_1988,grosberg_1993}, the loop-extrusion model \cite{marko_brahmachari_2019}, or a combination of the fractal-globule model with added crosslinks \cite{fgm_2023,subramanian_2023}.  

For force dipoles, we also discovered two other distinct behaviors that we consider fundamentally significant. Firstly, we observed the collapse of the fractal to a compact object when the force amplitude $f_0$ exceeds a critical value, for which the dynamical persistence length, generalized to account for a harmonic potential, exceeds the equilibrium distance between beads. This is reminiscent of the previously studied ``motility-induced phase transition" and other dynamical transitions in active systems \cite{cates_2015_review,phase_gonnella_2015,mips_2018,misp_2020,mips_soft_2021,mips_2022,dynamical_transition_cates_2019}. However, it should be noted that the structure in the collapsed state will be modified if excluded volume interaction is included, presenting a low cutoff for the collapsed state size. We, therefore, did not delve into a detailed study of its structure and dynamics.

Secondly, below the critical force mentioned above, we observed a slow rotational, ``crawling" motion of the whole fractal network, whose direction is random and chosen by the symmetry (isotropy) breaking associated with the random realization of the dipolar force locations. We have shown that this behavior is very similar to the one observed in triangular microswimmers for which the asymmetry is broken by choice \cite{swimmer_rizvi2018,rizvi_2020}. Even though the torque that each dipole exerts is vanishing, and so is the total torque, such a ``crawling" rotational motion {\it is possible}, in general, for a deformable body \cite{rotational_shapere_1987,rotational_shapere_1989,rotational_shapere_1989geometry,rotational_raz_2008} but not for a rigid body. Thus, force dipoles, which are expected to exist in chromatin \cite{bruinsma_2014}, should cause slow rotational motion of the chromosome (as in Fig.~\ref{fig_dipole_multi}), provided that chromatin is free to rotate, e.g., when it is {\it not} anchored to the nuclear envelope in Eukaryotic cells by lamin A (i.e., for lamin A depletion) or other linkers. Interestingly, a large-scale coherent motion has been experimentally observed in chromatin (possibly presenting local rotational motion of chromosomal territories), and backed up by simulations that include dipolar forces \cite{david_2018_coherence}. Rotational motion has been also observed in cytoplasmic flow and proposed -- conceptually in accord with our simulations -- to be caused by actomyosin force dipoles \cite{woodhouse_2012_circulation,cytoplasimic_flow_suzuki_2017}.

While our simulations were performed on a specific deterministic fractal, the Sierpinki gasket, our analytical theory, which agrees with the simulations, applies to any fractal, be it a deterministic or a disordered fractal, which makes our conclusions universal. We do not imply by any means that chromatin appears like the Sierpinski gasket; only that they are both fractal networks, which allows us to conclude about chromatin dynamics also from our simulations. Simulations of critical percolation clusters that form a disordered fractal below the percolation correlation length, are currently underway to further check the general applicability of our analytical theory and the universality of our conclusions. Further work is also planned to strengthen the connection between the observed chromatin dynamics in normal and treated cells (e.g., ATP-depleted and lamin A-depleted cells) and our fractal model \cite{weber_2012,lamina_a_garini_2015}.

\begin{acknowledgments}
 SS acknowledges the BGU Kreitmann School postdoctoral fellowship. We thank Sam Safran, 
 Itay Griniasty, Tom Witten, and Omer Granek for useful discussions, and the BGU Avram and Stella Goldstein-Goren fund for support.
\end{acknowledgments}

\section*{AUTHOR DECLARATIONS}
\subsection*{Conflict of Interest}
The authors have no conflicts of interest.
\subsection*{Author Contributions}
{\bf Sadhana Singh:} Software (lead); Visualization (lead); Methodology (equal); Formal analysis (equal); Writing – review \& editing (equal), Writing - original draft (equal). {\bf Rony Granek:} Conceptualization (lead); Methodology (equal); Formal analysis (equal); Writing – review \& editing (equal), Writing - original draft (equal).

\section*{Data Availability Statement}
The data that support the findings of this study are available from the corresponding author upon reasonable
request.

\appendix
\section{Model definition and analytical solution}\label{append_sec_analytical}
For the scalar elasticity Hamiltonian given by Eq.~(\ref{hamiltonian}),
the eigenstates (i.e., normal modes), $\Psi_{\alpha}$, are solutions of the eigenvalue equation
\beq
\omega_0^2 \sum_{j\in i }\l[\Psi_{\alpha,j}-\Psi_{\alpha,i}\r]=-\omega_{\alpha}^{2}\Psi_{\alpha,i}.
\lb{eigen}
\eneq
where $\{\omega_{\alpha}\}$ are the corresponding eigen-frequencies, $\Psi_{\alpha,i}$ is the $i$-th entry of of normal mode $\alpha$, and $j\in i$ denotes beads connected by springs to bead $i$. On a fractal network, these eigenstates are known as ``fractons", the vibrational excitations of the fractal \cite{alexander_1982,rammal_1983,alexander_1989,stauffer_1992,bunde_1997,granek_2005}. The eigenstates $\Psi_{\alpha}$ form an orthonormal set \cite{alexander_1989} such that 
$\sum_i \Psi_{\alpha,i}\Psi_{\beta,i}=\delta_{\alpha,\beta}$, and
$\sum_{\alpha} \Psi_{\alpha,i}\Psi_{\alpha,j}=\delta_{i,j}$. 
This allows to define a normal mode transform 
$\vec u_{\alpha} = \sum_{i}\vec u_i~\Psi_{\alpha,i}$, and an inverse transform 
$\vec u_i = \sum_{\alpha}\vec u_{\alpha}~\Psi_{\alpha,i}$.

On a fractal, the normal modes $\Psi_{\alpha}$ are strongly localized in topological space, unlike the oscillatory behavior characteristic of uniform networks. A disorder-averaged eigenstate may be defined according to
\beq
\bar\Psi(\omega_{\alpha},|\vec{\ell}_i-\vec{\ell}_j|)=N\langle\Psi_{\alpha,i}\Psi_{\alpha,j}\rangle_{\text{dis}} ~,
\label{append_dis_ave_eigen}
\eneq
where $\langle ...\rangle_{\text{dis}}$ denotes disorder averaging, i.e., averaging over all realizations of the fractal, keeping the bead $i$ and $j$ fixed, or averaging, within a given realization, overall different pair of beads $i$ and $j$ that have the same topological space distance $|\vec{\ell}_i-\vec{\ell}_j|$. 
($\vec{\ell}_i$ is the dimensionless coordinate of bead $i$ in topological space.) Note that mode normalization implies $\langle\Psi_{\alpha,i}^2\rangle_{\text{dis}}=1/N$.

It has been shown that, in topological space, $\bar\Psi(\omega_{\alpha},\ell)$ obeys the following scaling form \cite{stauffer_1992,alexander_1982,alexander_1989} 
\beq 
 \bar\Psi(\omega_{\alpha},l)=f\l[\l(\frac{\omega_{\alpha}}{\omega_o}\r)^{\frac{d_s}{d_l}}l\r],
\label{Psi}
\eneq 
where $f(y)$ is the scaling function. For $y\gg 1$, $f(y)$ is exponentially decaying, and, for
$y\ll 1$, $f(y)\simeq 1-C_0\times y^2$ where $C_0$ is a numerical constant \cite{alexander_1989,bunde_1992,bunde_1997}.

To analyze the dynamics of an active fractal network, we follow the Langevin equations described in Refs. \cite{granek_2005,granek_2011,shlomi_2012_prl,shlomi_2012_pre} for a passive fractal in the high damping limit, generalized for the active forces. In addition to the white thermal noise, beads experience stochastic active forces distributed randomly but uniformly over the network. The stochastic active forces are assumed to fluctuate independently of each other and to follow the random telegraph (on-off) process \cite{gardiner_1991}, where the random variables take two values, $0$ and $f_0$, describing the 'off' and 'on' states of the force. The auto-correlation function of the forces thus follows
\beq
\langle \vec f _i(t)\cdot \vec f _j(0)\rangle =f_0^2p^2\hat f_i\cdot\hat f_j+f_0^2\delta_{ij}p(1-p)\e^{-t/\tau},
\lb{append_force_corr-2}
\eneq
where $t$ is lag-time, $p$ is the probability for an 'on' state, and $\tau$ is the force correlation time.

We assume that the force directions are isotropically distributed. An isotropic distribution of $\hat f_i$'s implies 
\beq
\sum_{i,j,i\neq j}\hat f_i\cdot\hat f_j=0
\eneq
in an infinite network; however, in a finite system this sum may have a residual (random) value. This sum will nevertheless vanish on a finite system if we perform in addition disorder averaging, in which the force field spans over all its possible realizations.

For monopoles, the directional distribution is isotropic by definition. In the case of dipoles, although they are taken to align with the bonds, their overall directional distribution remains (statistically) isotropic due to the random realization of their locations. For off-lattice disordered fractals, such as randomly branched polymers, this directional distribution is effectively continuous. For on-lattice disordered fractals, such as critical percolation clusters, or deterministic fractals, such as the Sierpinski gasket used for the simulations, the distribution is isotropic within the allowed discrete directions (three, in the case of Sierpinski). Note that the spatial and directional correlations between any two forces making a local dipole are already accounted for in Eq.\ (\ref{dipole}).

For convenience, we consider the force fluctuations about their mean $f_0 p\hat f _i$. Introducing
\begin{equation}
\delta\vec f _i(t)=\vec f _i(t)-f_0 p\hat f _i~,
\end{equation}
we have
\beq
\langle \delta\vec f _i(t)\cdot \delta\vec f _j(0)\rangle =f_0^2\delta_{ij}p(1-p)\e^{-t/\tau}.
\lb{append_force_corr}
\eneq
It follows that $\{\vec u_{i}\}$ will stand henceforth for the displacements in bead positions about the mean coordinates, as dictated by the mechanical balance between the mean forces and the network elasticity.

For a Rouse-type fractal network, the set of Langevin equations of motion (same as Eq.(\ref{Langevin1})) of an active system are
\begin{equation}
\gamma \frac{d\vec u_i(t)}{dt} =m\omega_o^2\sum_{j\in i} \left(\vec{u}_j-\vec{u}_i\right) +\vec{\zeta}_i(t)+\vec f_i(t)\;, 
\label{append_Langevin1}
\end{equation}
where $\gamma$ is the local friction coefficient ($\gamma=3\pi\eta b$ in case of Stokes drag, where $\eta$ is the solvent viscosity and $b$ is the bead diameter), $\vec{\zeta}_i$ is the thermal white noise obeying FDT \cite{kubo_1966}, and $\vec{f}_i$ is the active force acting on $i$-th bead. 

Using the normal modes of the fractal network ${\Psi_{\alpha}}$, whose corresponding eigen-frequencies are ${\omega_{\alpha}}$, and transforming Eq.~(\ref{append_Langevin1}) to the normal mode space, we obtain
\begin{equation}
\frac{d\vec u_{\alpha}}{dt} = -\Gamma_{\alpha}\vec u_{\alpha} + \vec{\zeta}_{\alpha}(t) + \Lambda_{\alpha}\vec{F}_{\alpha}(t),
\label{append_Langevin4}
\end{equation}
Here $\vec u_{\alpha}(t)$ is the amplitude of a normal mode $\alpha$ at time $t$, $\Gamma_{\alpha}=m\omega_{\alpha}^2\Lambda_{\alpha}$ is the mode relaxation rate, where $\Lambda_{\alpha}$ is the mode mobility coefficient, $\Lambda_{\alpha}=1/\gamma$. 
$\vec{\zeta}_{\alpha}(t)$ is the (mode transformed) thermal white noise that obeys the fluctuation-dissipation theorem
\begin{equation}
\langle\vec{\zeta}_{\alpha}(t)\vec{\zeta}_{\beta}(t^{\prime})\rangle=2
k_BT\Lambda_{\alpha}\delta_{\alpha,\beta}\delta(t-t^{\prime}),
\lb{FD-theorem}\end{equation}
and, the last term, $\vec{F}_{\alpha}(t)$ is the mode transformed active noise fluctuations (about their mean), which takes the form: (i) for force monopoles
\beq
\vec{F}_{\alpha}(t)=\sum_j \delta\vec f_j(t) \Psi_{\alpha}(\vec r_j),
\eneq
and (ii) for force dipoles
\beq
\vec{F}_{\alpha}(t)=\sum_j \delta\vec f_j(t) \left[\Psi_{\alpha}(\vec r_j)-\Psi_{\alpha}(\vec r_j+\vec\epsilon_j)\right]
\eneq
where $\Psi_{\alpha}(\vec r_i)\equiv \Psi_{\alpha,i}$.

\subsection{Force auto-correlation function in mode space}\label{append_steady} 
We first calculate the auto-correlation function of the active force field in mode space:\\
(i) for force monopoles
\beq
\langle\vec{F}_{\alpha}(t)\vec{F}_{\alpha}(0)\rangle=\sum_{i,j}\langle\delta\vec f_i(t)\cdot \delta\vec f_j(0)\rangle \langle \Psi_{\alpha}(\vec r_i)\Psi_{\alpha}(\vec r_j)\rangle_{\text{dis}}~,
\label{append_force_mono_corr}
\eneq
(ii) for dipoles
\begin{align}
& \langle\vec{F}_{\alpha}(t)\vec{F}_{\alpha}(0)\rangle= \sum_{i,j} \langle\delta\vec f_i(t)\cdot \delta\vec f_j(0)\rangle\nonumber\times \nonumber \\
& [\langle \Psi_{\alpha}(\vec r_i)\Psi_{\alpha}(\vec r_j) \rangle_{\text{dis}} -\langle \Psi_{\alpha}(\vec r_i)\Psi_{\alpha}(\vec r_j+\vec\epsilon_j) \rangle_{\text{dis}}- \nonumber\\ 
& \langle \Psi_{\alpha}(\vec r_i+\vec\epsilon_i)\Psi_{\alpha}(\vec r_j) \rangle_{\text{dis}}+\langle \Psi_{\alpha}(\vec r_i+\vec\epsilon_i)\Psi_{\alpha}(\vec r_j+\vec\epsilon_j) \rangle_{\text{dis}}]
\label{append_force_dipoles_corr}
\end{align}
The averages in Eqs. (\ref{append_force_mono_corr})-(\ref{append_force_dipoles_corr}) is performed over both the stochastic evolution of the active forces and disorder:\\
(i) For force monopoles, we obtain, after disorder averaging, using Eq.~(\ref{append_dis_ave_eigen}),
\beq
\langle\vec{F}_{\alpha}(t)\vec{F}_{\alpha}(0)\rangle=\frac{1}{N}\sum_{i,j}\langle\delta\vec f_i(t)\cdot \delta\vec f_j(0)\rangle \bar\Psi(\omega_{\alpha},|\vec r_j-\vec r_i|),
\label{append_F_corr_mono}\eneq
and after ensemble (or time) averaging over the stochastic forces using Eq.~(\ref{append_force_corr}), and implementation of the normalization $\bar\Psi(\omega_{\alpha},0)=1$, we get
\beq
\langle\vec{F}_{\alpha}(t)\vec{F}_{\alpha}(0)\rangle =\phi f_0^2p(1-p)\e^{-t/\tau},
\label{append_f_corr_mono_1}
\eneq
where $\phi$ is the fraction of force monopoles. Note that Eq.~(\ref{append_f_corr_mono_1}) is
independent of the mode index $\alpha$.\\
(ii) For force dipoles, we first use Eq.~(\ref{append_dis_ave_eigen}) 
\begin{align}
& \langle\vec{F}_{\alpha}(t)\vec{F}_{\alpha}(0)\rangle=
\frac{1}{N}\sum_{i,j} \langle\delta\vec f_i(t)\cdot \delta\vec f_j(0)\rangle\nonumber\times \\
& [\bar\Psi(\omega_{\alpha},|\vec r_j-\vec r_i|) - \bar\Psi(\omega_{\alpha},|\vec r_j+\vec\epsilon_j-\vec r_i|)- \nonumber \\
& \bar\Psi(\omega_{\alpha},|\vec r_j-\vec r_i-\vec\epsilon_i|)+\bar\Psi(\omega_{\alpha},|\vec r_j-\vec r_i+\vec\epsilon_j-\vec\epsilon_i|)],
\end{align}
and after ensemble (or time) averaging over the stochastic forces, using Eq.~(\ref{append_force_corr}),
\beq
\langle\vec{F}_{\alpha}(t)\vec{F}_{\alpha}(0)\rangle = 2\phi f_0^2\left[\bar\Psi(\omega_{\alpha},0)-\bar\Psi(\omega_{\alpha},\epsilon)\right]p(1-p)\e^{-t/\tau},
\eneq
where $\phi$ is now the fraction of force dipoles, leading to 
\beq
\langle\vec{F}_{\alpha}(t)\vec{F}_{\alpha}(0)\rangle = 2\phi f_0^2\left[1-\bar\Psi(\omega_{\alpha},\epsilon)\right]p(1-p)\e^{-t/\tau}.
\eneq
Note that transforming Eq.~(\ref{Psi}) from topological space to real space, by using $l\sim (r/b)^{d_f/d_l}$, (where $d_f$ and $d_l$ are the fractal and topological dimensions, respectively), implies
\beq 
\bar\Psi(\omega_{\alpha},r)=f\l[(\omega_{\alpha}/\omega_o)^{d_s/d_l} (r/b)^{d_f/d_l}\r].
\eneq
Using $\epsilon=b$, and the low-frequency expansion for the eigenstates, we have $\bar\Psi(\omega_{\alpha},\epsilon)\simeq 1- C_0 (\omega_{\alpha}/\omega_o)^{2d_s/d_l}$, leading to
\begin{equation}
\langle\vec{F}_{\alpha}(t)\vec{F}_{\alpha}(0)\rangle \simeq 2C_0(\omega_{\alpha}/\omega_o)^{2d_s/d_l}\phi f_0^2p(1-p)\e^{-t/\tau}.
\end{equation}
To summarize, for both types of active forces, we have
\beq
\langle\vec{F}_{\alpha}(t)\vec{F}_{\alpha}(0)\rangle \simeq W_{\alpha}\phi f_0^2 p(1-p)\e^{-t/\tau},
\eneq
where
\beq
W_{\alpha}=Q\, \omega_{\alpha}^{\xi}
\eneq
such that
\beq
Q=\begin{cases} 
    1 & \text{for force monopoles},\\  
     2C_0 \omega_o^{-2d_s/d_l}  & \text{for force dipoles},
     \end{cases}
\eneq
and
\beq
\xi=\begin{cases}
    0 & \text{for force monopoles},\\
    2d_s/d_l & \text{for force dipoles.}
\end{cases}
\eneq

\subsection{Normal mode amplitude auto-correlation function}\label{append_amplitude} 
Solving the Langevin equation (\ref{append_Langevin4}), we obtain the auto-correlation function of the displacement at a steady state such that the absolute times $t_1$, $t_2$ are taken to infinity (or that active forces started to act at time $-\infty$), thereby all transient behavior has completely relaxed. The time difference $t=|t_2-t_1|$ is finite. Hence 
$\langle \vec{u}_{\alpha}(t)\cdot\vec{u}_{\alpha}(0)\rangle\equiv \langle \vec{u}_{\alpha}(t_2)\cdot\vec{u}_{\alpha}(t_1)\rangle$. We obtain
\begin{align}
\langle \vec{u}_{\alpha}(t)\cdot\vec{u}_{\alpha}(0) \rangle = & \frac{dk_BT}{m \omega_{\alpha}^2}\e^{-\Gamma_{\alpha}t} + \frac{\phi f_0^2p(1-p) W_{\alpha}\Lambda_{\alpha}^2}{\Gamma_{\alpha}\left(\Gamma_{\alpha}^2-\tau^{-2}\right)}\times \nonumber \\ 
& \left[\Gamma_{\alpha}\e^{-t/\tau}  -\tau^{-1}\e^{-\Gamma_{\alpha}t}\right],
\label{append_mode_corr}
\end{align}
and mean square normal mode amplitude is
\beq
\langle \vec{u}_{\alpha}^{\,2} \rangle =\frac{dk_BT}{m \omega_{\alpha}^2} + \frac{\phi f_0^2p(1-p) W_{\alpha}\Lambda_{\alpha}^2}{\Gamma_{\alpha}\left(\Gamma_{\alpha}+\tau^{-1}\right)}.
\label{append_mode_mean_squre}
\eneq
In Eq.~(\ref{append_mode_mean_squre}), the first term is the thermal contribution and the second is the contribution of the stochastic force fluctuations. For $\tau\to\infty$ the latter reduces to
\beq
\langle \vec{u}_{\alpha}^{\,2} \rangle =\frac{dk_BT}{m \omega_{\alpha}^2} + \frac{\phi f_0^2p(1-p) W_{\alpha}} {(m\omega_{\alpha}^2)^2}.
\eneq

\subsection{Mean square displacement}\label{append_msd}
Expanding the two-point correlation function \cite{shlomi_2012_pre} in terms of the exact normal modes $\Psi_{\alpha}(\vec{r})$, and performing disorder averaging, we find the required MSD as the sum of a purely thermal contribution, $\langle\Delta\vec{r}(t)^2\rangle_{\text{th}}$, and a purely active contribution, $\langle\Delta\vec{r}(t)^2\rangle_{\text{ac}}$:
\begin{align}
\langle\Delta\vec{r}(t)^2\rangle & \equiv \langle \left(\vec{r}_i(t)-\vec{r}_i(0) \right)^{2} =\langle \left(\vec{u}_i(t)-\vec{u}_i(0) \right)^{2} \rangle \nonumber \\ 
& = \frac{2}{N}\sum_{\alpha} \left(\langle \vec{u}_{\alpha}^{\,2} \rangle - \langle \vec{u}_{\alpha}(t)\cdot \vec{u}_{\alpha}(0) \rangle \right),
\label{append_MSDgeneral}
\end{align}
using the result Eq.~(\ref{append_mode_corr}), and (\ref{append_mode_mean_squre}) we thus obtain
\beq
\langle\Delta\vec{r}(t)^2\rangle = \langle\Delta\vec{r}(t)^2\rangle_{\text{th}}+\langle\Delta\vec{r}(t)^2\rangle_{\text{ac}},
\label{append_MSD_total}
\eneq
where
\begin{align}
& \langle\Delta\vec{r}(t)^2\rangle_{\text{th}} = \frac{1}{N}\sum_{\alpha} \frac{2dk_BT}{m \omega_{\alpha}^2}\left(1-\e^{-\Gamma_{\alpha}t} \right)
\label{append_MSD_total_th}\\
& \langle\Delta\vec{r}(t)^2\rangle_{\text{ac}} = \frac{1}{N}\sum_{\alpha}\frac{2\phi f_0^2p(1-p) W_{\alpha}\Lambda_{\alpha}^2}{\Gamma_{\alpha}\left(\Gamma_{\alpha}+\tau^{-1}\right)}\times \nonumber \\
& \left(1 + \frac{\tau^{-1}}{\Gamma_{\alpha}-\tau^{-1}}\e^{-\Gamma_{\alpha}t} -\frac{\Gamma_{\alpha}}{\Gamma_{\alpha}-\tau^{-1}}\e^{-t/\tau} \right).
\label{append_MSD_total_ac}
\end{align}
In Eqs.\ (\ref{append_MSD_total_th})-(\ref{append_MSD_total_ac}), the term corresponding to $\omega_{\alpha}=0$ describes the translational motion and should be omitted from the sum. Nevertheless, the translational CM motion can be deduced analytically by taking the limit $\omega_{\alpha}\to 0$. It follows that both thermal and active CM motions are diffusive; the thermal CM motion is
\beq
\langle\Delta\vec R_{CM}(t)^2\rangle_{\text{th}}=\frac{2dk_BT}{N\gamma}t,
\label{th_CM_motion}
\eneq
and the active CM motion is
\beq
\langle\Delta\vec R_{CM}(t)^2\rangle_{\text{ac}}=\begin{cases} 
    \frac{2\phi f_0^2 p(1-p)\tau}{N\gamma^2}t & \text{for force monopoles},\\  
     0  & \text{for force dipoles},
     \end{cases}
\label{ac_CM_motion}
\eneq

Using the vibrational DOS $g(\omega)=n_o\omega^{d_s-1}$, where $n_o=N d_s/\omega_o^{d_s}$ is 
chosen such that $\int_0^{\omega_o} d\omega g(\omega)=N$, we can approximate the sum in 
Eqs.~(\ref{append_MSD_total_th})-(\ref{append_MSD_total_ac}) to an integral over the frequency. 
Note that the lower and upper integration limits set the shortest 
$\tau_0=\Gamma(\omega_{0})^{-1}$ and longest relaxation times 
$\tau_N=\Gamma(\omega_{\text{min}})^{-1}$ where 
$\omega_{\text{min}}\simeq \omega_o (R_g/b)^{-d_f/d_s}\simeq \omega_o N^{-1/d_s}$ of 
flexible network dynamics.
As already described in Refs. \cite{granek_2005,shlomi_2012_pre,shlomi_2012_prl,granek_2011}, 
focussing on the time regime $\tau_0\ll t\ll \tau(R_g)$, the thermal MSD exhibits subdiffusion
\beq
\langle\Delta\vec{r}(t)^2\rangle_{\text{th}}=B_{\text{th}}~t^{\nu_{\text{th}}},
\label{append_thermal_msd}
\eneq
with $\nu_{\text{th}}=1-\frac{d_s}{2}$, provided that $d_s<2$, and the amplitude (prefactor) 
\beq
B_{\text{th}}=2dC_{\text{th}}\frac{d_s}{\omega_0^{d_s}}\frac{k_BT}{ m^{\frac{d_s}{2}}}   \gamma^{\frac{d_s}{2}-1}.
\label{append_thermal_msd_Bth}
\eneq
Here, $C_{\text{th}}$ is a numerical prefactor, $C_{\text{th}}=\Gamma[\frac{d_s}{2}]/(2-d_s)$ in the Rouse model ($\Gamma[x]$ is the Gamma function).

An analytic evaluation of the other active origin terms is partly possible for a large system such that the frequency spectrum is dense. The integral approximation of Eq.~(\ref{append_MSD_total_ac}) is
\begin{align}
& \langle\Delta\vec{r}(t)^2\rangle_{\text{ac}} =\frac{2d_s}{\omega_o^{d_s}}\phi f_0^2p(1-p)\times \nonumber\\ 
& \int_{\omega_{\text{min}}}^{\omega_o} d\omega\omega^{d_s-1}\frac{ W(\omega)\Lambda(\omega)^2}{\Gamma(\omega)\left(\Gamma(\omega)+\tau^{-1}\right)}\times \nonumber \\ 
& \left(1 + \frac{\tau^{-1}}{\Gamma(\omega)-\tau^{-1}}\e^{-\Gamma(\omega)t} -  \frac{\Gamma(\omega)}{\Gamma(\omega)-\tau^{-1}}\e^{-t/\tau}\right).
\label{append_two_point_integral}
\end{align}

Focusing on the intermediate time regime $\tau_0\ll t\ll \tau_N$, and times shorter than the force correlation time such that $\tau_0\ll t\ll\tau\ll \tau_N$, we find superdiffusive, ballistic-type behavior $\langle\Delta\vec{r}(t)^2\rangle_{\text{ac}}\sim t^2$ for both force types, monopoles and dipoles. More precisely, we obtain
\beq
\langle\Delta\vec{r}(t)^2\rangle_{\text{ac}}\simeq B_{\text{acs}} t^2\;\;\;\;\;~~~~~~t\ll \tau,
\eneq
where 
\beq
B_{\text{acs}} = \phi f_0^2p(1-p)\frac{d_s}{\omega_o^{d_s} }\int_{\omega_{\text{min}}}^{\omega_0} d\omega 
\; \omega^{d_s-1} \frac{W(\omega)\Lambda(\omega)^2\tau^{-1}}{\Gamma(\omega)+\tau^{-1} }\;.
\label{append_two_point_integral-1}
\eneq
Setting the lower and upper limits of integration in Eq.~(\ref{append_two_point_integral-1}) to $0$ and infinity, respectively, and performing the integration, we find, {\it provided that} $1-d_s/2-\xi/2>0$,
\beq
B_{\text{acs}} = C_{\text{acs}}\,Q\, p(1-p) \frac{d_s}{\omega_o^{d_s}} \frac{\gamma^{\frac{d_s+\xi-4}{2}}}{m^{\frac{d_s+\xi}{2}}}\phi f_0^2 \tau^{-\frac{d_s+\xi}{2}}, 
\eneq
where $C_{\text{acs}}$ is a numerical factor, $C_{\text{acs}}=\int_{0}^{\infty}d \omega \frac{\omega^{d_s+\xi-1}}{\omega^2+1}= \frac{1}{2}B\left(\frac{d_s+\xi}{2},1-\frac{d_s+\xi}{2}\right)$, and $B(x,y)$ is the Beta function \cite{gradshteyn_2014}. For force monopoles, this implies
\beq
B_{\text{acs}} = C_{\text{acs}} p(1-p) \frac{d_s}{\omega_o^{d_s}} \frac{\gamma^{\frac{d_s}{2}-2}}{m^{\frac{d_s}{2}}}\phi f_0^2 \tau^{-\frac{d_s}{2}} 
\eneq
with $C_{\text{acs}}=\frac{1}{2}B\left(\frac{d_s}{2},1-\frac{d_s}{2}\right)$, hence $B_{\text{acs}}\propto \phi f_0^2 \tau^{-d_s/2}$. In the case of force dipoles, we note that the condition $1-d_s/2-d_s/d_l>0$ is commonly {\it not} obeyed in familiar fractals; in the rare situation where it is obeyed $B_{\text{acs}}\propto \phi f_0^2\tau^{-d_s/2-d_s/d_l}$ for force dipoles. Otherwise, for $1-d_s/2-d_s/d_l<0$ the integral in Eq.~(\ref{append_two_point_integral-1}) diverges with the upper limit $\omega_0$, and we find
\beq
B_{\text{acs}} = \frac{d_s}{d_s+\xi-2}Q\,p(1-p)\,\omega_o^{\frac{2d_s}{d_l}-2}m^{-1}\gamma^{-1}\phi f_0^2 \tau^{-1} 
\eneq
such that $B_{\text{acs}}\propto \phi f_0^2 \tau^{-1}$. Note that the influence of $\tau$ at times $t\ll \tau$ does not contradict causality, as we describe the fluctuations at a steady state where the laboratory time ($t_{tot}$, c.f. Eq.~(\ref{append_time_ave_msd})) has surpassed $\tau$ many times.

For times much longer than the force correlation time, $t\gg\tau$, Eq.~(\ref{append_two_point_integral}) can be approximated as
\begin{align}
& \langle\Delta\vec{r}(t)^2\rangle_{\text{ac}} \simeq 2\phi f_0^2p(1-p)\tau \frac{d_s}{\omega_o^{d_s}} \times \nonumber \\ 
& \int_{\omega_{\text{min}}}^{\omega_0} d\omega~\omega^{d_s-1}\frac{W(\omega)\Lambda(\omega)^2}{\Gamma(\omega)}\left(1 - \e^{-\Gamma(\omega)t} \right)\;.
\label{append_two_point_integral_1}
\end{align}
{\it Provided that} $1-d_s/2-\xi/2>0$, setting the lower and upper limits of integration to $0$ and infinity (respectively) and scaling of the integral leads to
\beq
\langle\Delta\vec{r}(t)^2\rangle_{\text{ac}}= B_{\text{ac}}t^{\nu_{\text{ac}}},
\eneq
where $\nu_{\text{ac}}=1-d_s/2-\xi/2$, and
\beq
B_{\text{ac}}= 2C_{\text{ac}}~Qp(1-p)\frac{d_s}{\omega_0^{d_s}}\frac{\gamma^{\frac{d_s+\xi-4}{2}}}{m^{\frac{d_s+\xi}{2}}}\phi f_0^2\tau\; . 
\eneq
Here $C_{\text{ac}}$ is a numerical constant, $C_{\text{ac}}=\int_0^{\infty}d\omega~\omega^{d_s+\xi-3}(1-\e^{-\omega^2})=\Gamma(\frac{d_s+\xi}{2})/(2-d_s-\xi)$ where $\Gamma(x)$ is Gamma function \cite{gradshteyn_2014}. Thus, for force monopoles, $\nu_{\text{ac}}=1-d_s/2$, that is identical to the thermal subdiffusion exponent, and
\beq
B_{\text{ac}}= 2C_{\text{ac}}~p(1-p)\frac{d_s}{\omega_0^{d_s}}\frac{\gamma^{\frac{d_s}{2}-2}}{m^{\frac{d_s}{2}}}\phi f_0^2\tau\; ;
\eneq
hence, the ratio of the active to thermal MSD amplitudes is
\beq
\frac{B_{\text{ac}}}{B_{\text{th}}} = p(1-p)\frac{\phi f_0^2\tau}{dk_BT \gamma}.
\eneq
For force dipoles, since, as noted usually $1-d_s/2-d_s/d_l<0$, the integral in Eq.~(\ref{append_two_point_integral_1}) diverges with the upper bound. Therefore, an anomalous subdiffusion regime is absent, and the MSD crosses over rapidly to a constant value (see below).

\subsection{Static mean square displacement}\label{append_static_msd} 
For $t\gg\{\tau_N,\tau\}$ the MSD saturates at
\beq
\langle\Delta\vec{r}(\infty)^2\rangle=2\langle\vec{r}^2\rangle,
\eneq
in which the thermal and active contributions are also additive,
\beq
\langle\vec{r}^2\rangle=\langle\vec{r}^2\rangle_{\text{th}}+\langle\vec{r}^2\rangle_{\text{ac}}~.
\eneq
As previously shown for a thermal (passive) fractal network with $d_s<2$, the thermal contribution exhibits the generalized Landau-Peierls instability $\langle \vec{r}^2\rangle_{\text{th}}\sim N^{\frac{2}{d_s}-1}$ \cite{burioni_2002,burioni_2004}. The complete expression for static thermal MSD is 
\beq
\langle\vec{r}^2\rangle_{\text{th}}= \frac{d_s}{2-d_s} \frac{d k_BT}{m\omega_o^{2}}N^{\frac{2}{d_s}-1}.
\label{append_thermal-static-LP}\eneq
For the active contribution to the static MSD, we use the second term in Eq.~(\ref{append_mode_mean_squre}) to give the mean square amplitude
\beq
\langle\vec{r}^2\rangle_{\text{ac}}=\phi f_0^2p(1-p) \frac{d_s}{\omega_o^{d_s}} \int_{\omega_{\text{min}}}^{\omega_0} d\omega\, \omega^{d_s-1} \frac{ W_{\alpha} \Lambda_{\alpha}^2}{\Gamma_{\alpha}\left(\Gamma_{\alpha}+\tau^{-1}\right)}~.
\label{append_MSD-static-active}
\eneq
The integral in Eq.~(\ref{append_MSD-static-active}) may diverge with the lower limit $\omega_{\text{min}}\to 0$ ($N\to\infty$), where $\omega_{\text{min}}\sim \omega_o N^{-1/d_s}$, giving rise to instability that resembles the Landau-Peirels instability. This divergence occurs if $1-d_s/2-\xi/2>0$, in which case
\beq
\langle\vec{r}^2\rangle_{\text{ac}}=\left(\frac{d_s}{2-d_s-\xi}\right)\frac{Q}{m\omega_o^{2-\xi} \gamma }~\phi f_0^2\tau  p(1-p) N^{\frac{2-\xi}{d_s}-1}.
\label{append_MSD-static-active-1}
\eneq
Hence, for force monopoles ($\xi=0$) and with $d_s<2$, this yields an {\it identical} divergence with increasing system size similar to the thermal Landau-Peierls instability,
\beq
\langle\vec{r}^2\rangle_{\text{ac}}=\left(\frac{d_s}{2-d_s}\right) \frac{\phi f_0^2\tau  p(1-p)}{m\omega_o^{2} \gamma }~ N^{\frac{2}{d_s}-1},
\label{append_MSD-static-active-2}
\eneq
i.e., $\langle\vec{r}^2\rangle_{\text{ac}}\sim N^{\frac{2}{d_s}-1}$. The active noise contribution will thus prevail, provided that the force magnitude and correlation time are sufficiently large. For an active system with force monopoles, we may thus assign {\it an effective temperature}
\beq
T_{\text{eff}}=T+ \frac{\phi f_0^2\tau p(1-p)}{dk_B\gamma}\;,
\eneq
that can faithfully describe, {\it both} the anomalous subdiffusion regime and the static (saturation) MSD, $2\langle\vec{r}^2\rangle$. 

Turning to force dipoles since, as already mentioned in typical fractals, we mostly have $1-d_s/2-d_s/d_l<0$, Eq.~(\ref{append_MSD-static-active-1}) does not apply, and $\langle\vec{r}^2\rangle_{\text{ac}}$ -- Eq.~(\ref{append_MSD-static-active}) -- diverges with the upper integration limit ($\omega_0$), becoming essentially independent of $N$. This implies that, surprisingly, $\langle\vec{r}^2\rangle_{\text{th}}$ {\it may dominate} the total static MSD for large enough systems due to the Landau-Peierls instability.

\section{Numerical evaluation of analytical results and Langevin dynamics 
simulations for Sierpinski gasket}\label{append_sec_numerical} 
\begin{figure}
\centering
\includegraphics[width=0.33\textwidth,trim={0.1cm 0.9cm 0.2cm 0.9cm},clip]{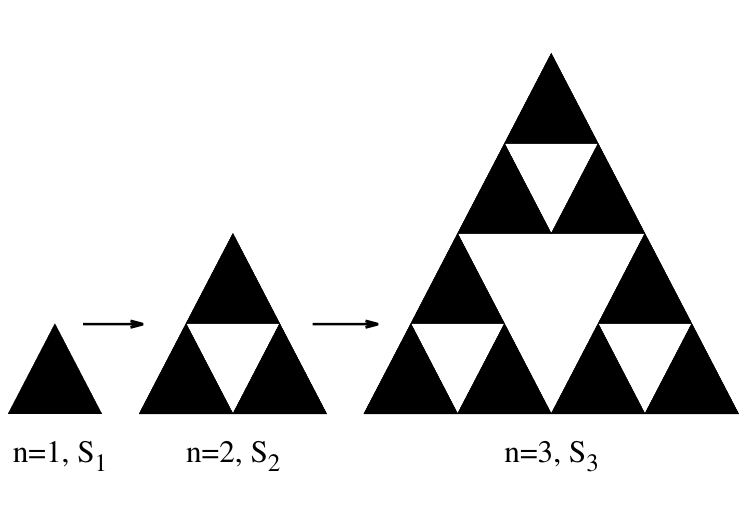}
\caption{{\bf The two-dimensional Sierpinski gasket} at different generations S$_n$, $n=1,2,3$. At generation $n=1$, Sierpinski gasket S$_1$ is an equilateral triangle with sides $a$. Generation $n+1$, S$_{n+1}$, is constructed by placing three $n_{th}$ generation structures S$_n$ adjacent to each other such that vertices of three triangles S$_n$ are touching each other with three common nodes as shown in the figure. The same process is iterated to get the next higher generation of Sierpinski gasket.}
\label{fig_schematic}
\end{figure}

Here we apply the general analytical theory for arbitrary fractals, Eqs.~(\ref{append_MSD_total_th}--\ref{append_MSD_total_ac}), and Langevin dynamics simulations, to the Sierpinski gasket (Fig.~\ref{fig_schematic}). We construct different generations of the Sierpinski gasket, in which the nodes (vertices) represent beads with identical masses $m$ that are connected by identical harmonic springs with spring constant $m\omega_0^2$ and equilibrium distance $b$, mimicking a bead-spring model of the fractal. We simulate the Sierpinski network dynamics under active force monopoles and dipoles, following Langevin dynamics simulations in the high damping limit. The set of Langevin equations of motion follow Eq.~(\ref{Langevin1}) with Hamiltonian $H=\frac{1}{2}m\omega_0^2\sum_{< ij >}(\vec{r}_i-\vec{r}_j-b \hat{r}_{ij} )^2$, where $\hat{r}_{ij}$ is the unit vector of distance $\vec{r}_i-\vec{r}_j$ between beads $i$ and $j$. For $b=0$, this interaction potential reduces to the scalar elasticity Hamiltonian (\ref{hamiltonian}) studied analytically above and in the main text. The simulations, therefore, will compare this more realistic model predictions to those used in the analytical calculation. 

\begin{figure}
\centering
\includegraphics[width=0.27\textwidth,trim={0.1cm 0.1cm 1.1cm 0.1cm},clip]{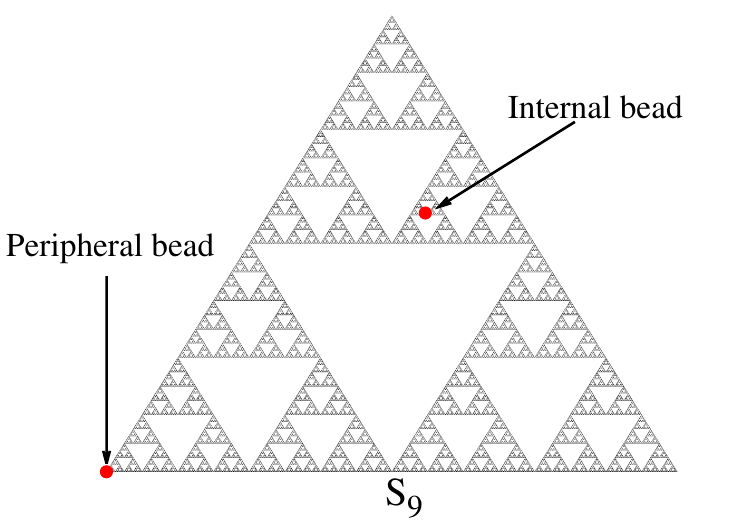}
\caption{A two-dimensional Sierpinski gasket S$_9$ of generation $n=9$. Here, the red mark shows the position of the internal and peripheral beads. For the MSD calculation of an arbitrary bead in numerical simulation, we have considered the position of the internal bead. 
}
\label{fig_gen_9}
\end{figure}

Without changing the notations (for simplicity), we transform to dimensionless variables as follows: $\vec{r}_i\to \vec{r}_i/b$, $\vec{f}_i\to \vec{f}_i/(m\omega_0^2 b)$ and $\vec{\zeta}_i\to \vec{\zeta}_i/(m\omega_0^2 b)$, and introduce a dimensionless time $t\to t/\tau_0$ where $\tau_0=\gamma/m\omega_0^2$ is the shortest relaxation time, such that the reduced  equation of motions are
\begin{equation}
\frac{d{\vec{r}}_i(t)}{dt}=-\sum_{j\in i}\left[\vec{r}_i-\vec{r}_j- \hat{r}_{ij} \right] + \vec{\zeta}_i(t) + \vec{f}_i(t), 
\label{gle1}
\end{equation}
 with a corresponding FD theorem
\beq
\langle \vec{\zeta}_i(t)\cdot\vec{\zeta}_j(t')\rangle=\frac{2dk_BT}{m\omega_0^2 b^2}\delta_{ij}\delta(t-t').
\eneq
Here, $d$ is the Euclidean embedding space dimension. In the simulations that follow, we have used $\frac{k_B T}{m\omega_0^2 b^2}=1$. For active force, we have taken various values of $f_0$, $\tau$, and $p=0.5$ throughout the simulations. 
The relative strength of the active force compared to the thermal force is expressed in dimensionless units as $\frac{f_0^2\tau p(1-p)}{d}$. 

To solve the equation of motions numerically, we have used the three-values Gear predictor-corrector algorithm \cite{allen.1987}. The integration step $\Delta t$ is $0.05\tau_0$. To achieve equilibrium (before data acquisition), we first run the simulations for times much longer (two to three orders of magnitude longer) than the longest relaxation $\tau_N$. After reaching a steady state, the simulations were run sufficiently long to calculate the time-averaged MSD accurately. The time-averaged MSD of an arbitrarily chosen bead (the internal bead shown in fig.~\ref{fig_gen_9}) of the fractal network is calculated using the following familiar expression
\beq
\langle (\vec{r}(t)-\vec{r}(0))^2\rangle = \frac{1}{t_{tot}-t}\sum_{t_i}^{t_{tot}-t} [\vec{r}(t_i+t)-\vec{r}(t_i)]^2,
\label{append_time_ave_msd}
\eneq 
where $t$ is the lag time, and $t_{tot}$ is the time window of data acquisition, obeying $t_{tot}\gg t$. In all figures presenting the analytical and simulation results, distances are in units of $b$, forces are in units of $m\omega_o^2 b$, and times are in units of $\tau_0=\gamma/m\omega_o^2$.

\subsection{Force monopoles}\label{append_sec_mono}
Force monopole directions are randomly picked from a uniform (isotropic) distribution. Thus, the total force on an infinite system ($N\to\infty$), and hence the drift velocity, would vanish. For a finite network, the average total force on the fractal depends on the actual realization of the forces. If we average on many realizations, the mean total force would again vanish, with a standard deviation $\sim \phi f_0 p \sqrt{N}$. Since the total friction is $N\gamma$, we estimate the typical magnitude of the drift velocity (whose direction is random) as $v_{\text{drift}}\sim \frac{\phi f_0 p}{\gamma \sqrt{N}}$. The effect of this drift velocity can be seen in Fig.~\ref{fig_mono_msd}(a) at long times.
To remove the effect of drift velocity, the expression for MSD is modified as following
\beq
\langle (\vec{r}(t)-\vec{r}(0))^2\rangle = \frac{1}{t_{\text{tot}}-t}\sum_{t_i}^{t_{\text{tot}}-t} (\vec{r}(t_i+t)-\vec{r}(t_i)-\langle \vec{r}(t)\rangle)^2),
\label{append_mono_time_ave_msd}
\eneq
where
$\langle (\vec{r}(t)\rangle = \frac{1}{t_{\text{tot}}-t}\sum_{t_i}^{t_{\text{tot}}-t} (\vec{r}(t_i+t)-\vec{r}(t_i))=\langle v_{\text{drift}}\rangle t$ is the mean displacement due to drift velocity.
\begin{figure}
\centering
\includegraphics[width=0.34\textwidth,trim={1.1cm 0cm 3.8cm 0cm},clip]{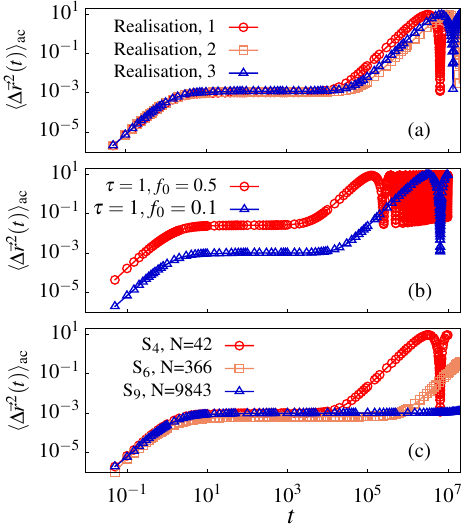}
\caption{{\bf Active network with force dipoles:} MSD of an internal bead of an active Sierpinski
  gasket S$_4$ (N=42): (a) MSD for three different realizations of the gasket, with $f_0=0.1$, $p=0.5$, and $\tau=1$. (b) MSD for two different force amplitudes, with $\tau=1$ and $p=0.5$.
  (c) MSD of three different generations of gasket for $f_0=0.1$, $p=0.5$, and $\tau=1$.
  In these figures, the crossover from constant to ballistic motion followed by oscillations
  is observed at long times, indicating a rotational motion with mean velocity $\frac{1}{N}\sum{\Omega_i}$, and the corresponding crossover time depends on this mean value. The different random realizations of force dipoles in (a), showing crossover at different times, indicate the random nature of the mean rotational velocity. Higher force amplitudes, as shown in (b), lead to an increase in rotational frequency, whereas higher generations of gaskets shown in (c) depict a crossover at longer times, suggesting a decrease in rotational frequency. (see also movie SM-1 to SM-3). $r$ is in unit of $b$, $f_0$ in unit of $m\omega_0^2 b$, and $t$ and $\tau$ are in units of $\tau_0=\gamma/(m\omega_0^2)$.}
 \label{fig_dipole_multi_si}
\end{figure}

\subsection{Force dipoles}\label{append_sec_dipole}
Many active soft-matter systems can produce force dipoles. They emerge in various biological systems such as ``actomyosin", i.e., an actin filament network in which myosin motors are dispersed \cite{drescher_2011}. These internal forces must satisfy the conditions of zero net momentum transfer and zero angular momentum transfer, such that the total force and torque vanish \cite{ramaswamy_2002_prl,ramaswamy_2010_anu_rev}. This implies that for each force acting on a network bead $i$, there is an equal and opposite force acting on another connected bead $j$.
\begin{equation}
\vec{F}_{tot,j}=-\vec{F}_{tot,i},
\end{equation}
\begin{equation}
\sum_i \vec{F}_{tot,i}=0,
\end{equation}
where $\vec{F}_{tot,i}=-\nabla_i H+\vec{f}_i$ is the total force acting on it $i$-th bead. Thus, there is no net force on the system, and so there is no CM translational motion. Furthermore, during the simulations, each dipole orientation remains parallel to bond vector $\vec{R}_i-\vec{R}_j$, where $\vec{R}_i$ is the position vector in the CM frame of reference. Each force pair contributes zero torque to the system; therefore, the net torque on the system vanishes too, i.e.
\begin{equation}
\vec{R}_i \times \vec{F}_{tot,i}+\vec{R}_j\times \vec{F}_{tot,j}=(\vec{R}_i-\vec{R}_j)\times \vec{F}_{tot,i}=0,
\end{equation}
\begin{equation}
\sum_i \vec{T}_{i} = 0,
\end{equation}
where $\vec{T}_i = \vec{R}_i \times \vec{F}_{tot,i}$ (`$\times$' stands for the vector product) is the
torque acting on $i$-th bead.
\begin{figure}
\centering
\includegraphics[width=0.31\textwidth,trim={0.8cm 0.1cm 1cm 0.6cm},clip]{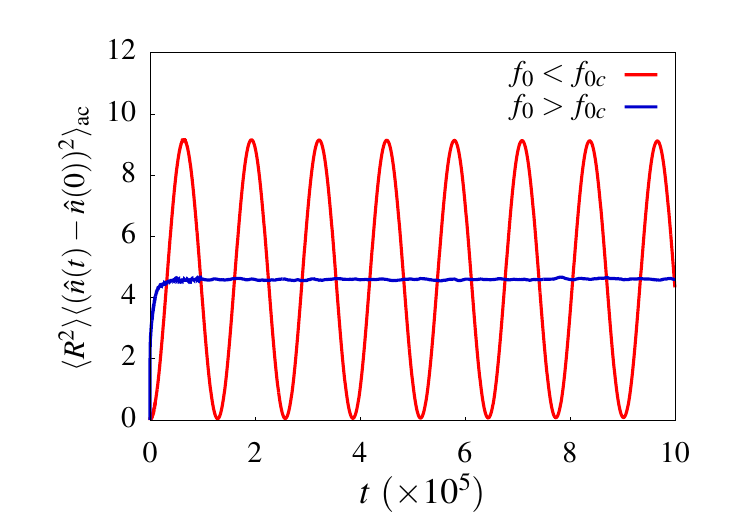}
\caption{{\bf Active network with force dipoles:} Mean square angular displacement (MSAD), $\langle R^2\rangle \langle(\hat n(t)-\hat n(0))\rangle_{ac}$, of an internal bead of an active Sierpinski gasket S$_4$ (N=42) subjected to force dipoles for force amplitude $f_0$ below and above the critical force amplitude $f_{0c}$ with $\tau=1$ and $p=0.5$. $r$ is in unit of $b$, $f_0$ in unit of $m\omega_0^2 b$, and $t$ and $\tau$ are in units of $\tau_0=\gamma/(m\omega_0^2)$.}
 \label{fig_dipole_MSAD_force}
\end{figure}

However, according to the equation of motion, each force is balanced by the friction force, hence
\begin{equation}
\vec{F}_{tot,i} = \gamma \vec{v}_i,
\label{eq_motion_total}
\end{equation}
where $\vec{v}_i$ is the velocity of $i$-th bead. Multiplying both sides of Eq.~(\ref{eq_motion_total}) by $\vec{R}_i$, we can get the torque acting on $i$-th bead,
\begin{equation}
\vec{T}_i =\vec{R}_i \times \vec{F}_{tot,i} = \gamma(\vec{R}_i \times \vec{v}_i),
\end{equation}
Using $\vec{v}_i=\vec{\Omega}_i \times \vec{R}_i$ where $\Omega_i$ is the angular velocity we obtain
\begin{equation}
 \vec{T}_i = \Gamma(\vec{R}_i \times (\vec{\Omega}_i \times \vec{R}_i)),
\end{equation}
\begin{equation}
\vec{R}_i \times (\vec{\Omega}_i \times \vec{R}_i)= \vec{R}_i(\vec{\Omega}_i\cdot\vec{R}_i) + \vec{\Omega}_i(\vec{R}_i\cdot\vec{R}_i),
\end{equation}
where
\begin{equation}
\vec{R}_i(\vec{\Omega}_i\cdot\vec{R}_i)=0~.
\end{equation}
The condition of net zero torque on the network, therefore, implies
\begin{equation}
\frac{1}{N}\sum_i \vec{\Omega}_i |\vec{R}_i|^2 =\overline{\vec{\Omega}|\vec{R}|^2} = 0,
\label{mean_ang_vel_r2}
\end{equation}
which does {\it not} necessarily implies
\begin{equation}
\frac{1}{N}\sum_i \vec{\Omega}_i =\overline{\vec{\Omega}}=0\;.
\label{mean_ang_vel}
\end{equation}
Hence, a ``crawling" (non-coherent) rotation of the whole object with mean velocity
$\overline{\vec{\Omega}}$ may exist under the action of active dipole forces; see Fig~\ref{fig_dipole_multi}(b), (c), and (d). The effect of this rotational motion can be seen in the MSD as a crossover from a constant to oscillatory behavior, whose initial presentation is ballistic-like, see Figs.~\ref{fig_dipole_multi_si}.

\begin{figure}
\centering
\hspace{-0.8cm}\includegraphics[width=0.15\textwidth,trim={0.3cm 6cm 5.8cm 7.6cm},clip]{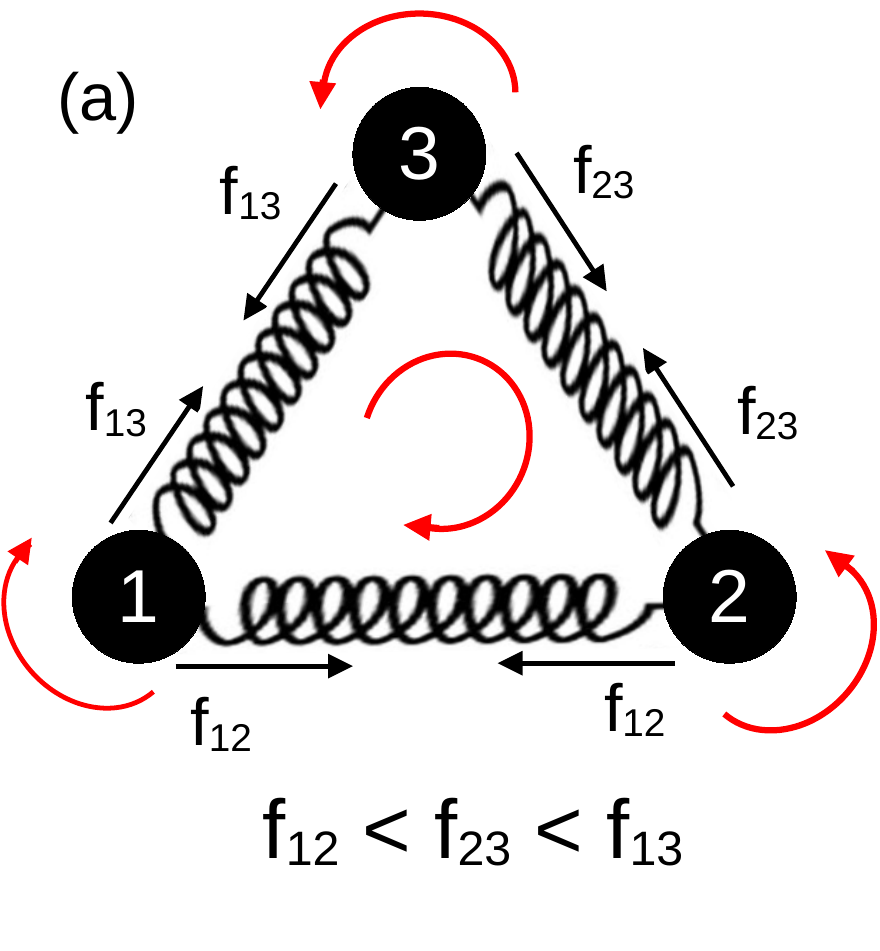}
\hspace{0.7cm}\includegraphics[width=0.15\textwidth,trim={0.3cm 6cm 5.8cm 7.6cm},clip]{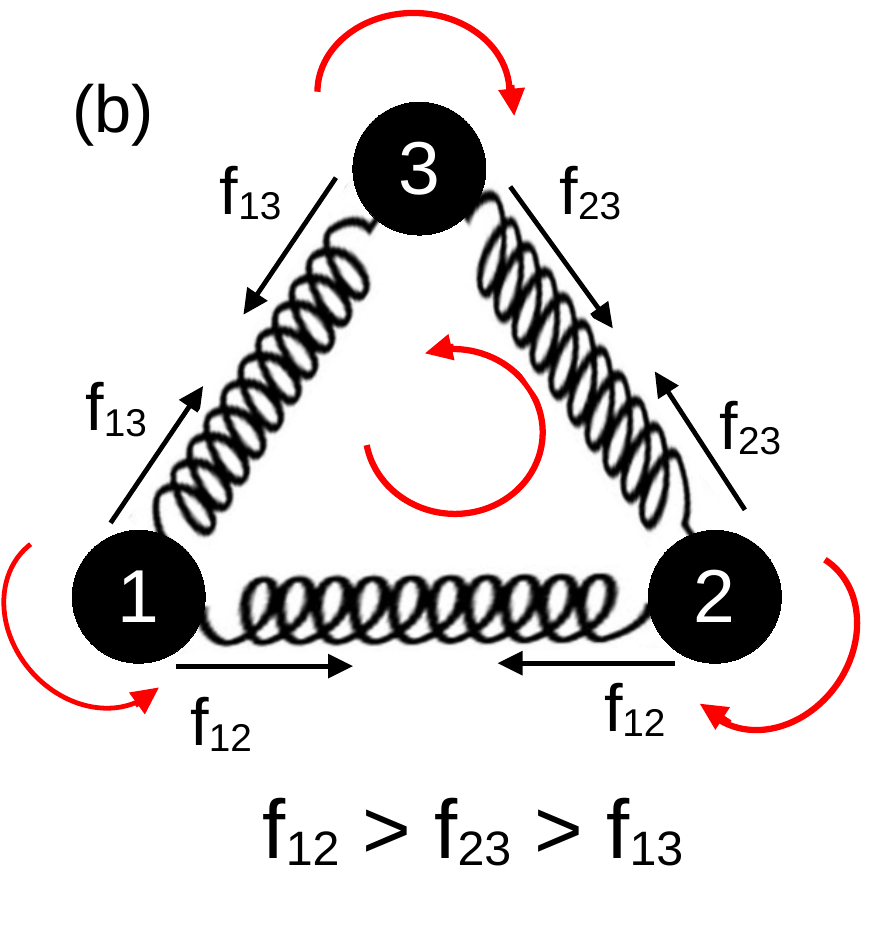}
\includegraphics[width=0.42\textwidth]{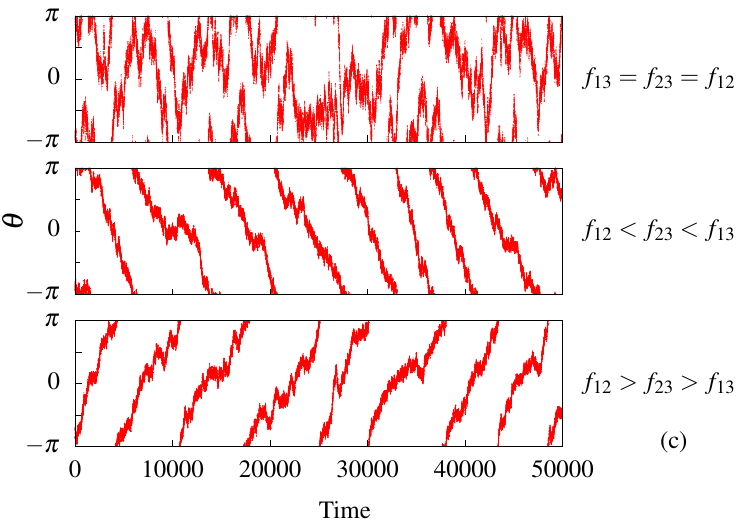}
\caption{{\bf Triangular micro-swimmer.} (a) and (b) Schematic of a triangle, the smallest generation unit of Sierpinski gasket, employed with force dipoles along each bond, with either clockwise decreasing or increasing strengths. (c) Time variation of the stochastic angular position, $\theta(t)$ (i.e., the angle between $\vec{R}(0)$ and $\vec{R}(t)$) of an arbitrary bead in the CM frame of reference with different order of forces (in the range of 0.1-0.5) along bonds for given $\tau=1$ and $p=0.5$. Notice the continuous periodic change in $\theta$ (even though local fluctuations are present) from $\pi$ to $-\pi$ and $-\pi$ to $\pi$, indicating anti-clockwise and clockwise rotations, respectively, with the direction of rotations depending on the order of asymmetric forces. Symmetric forces lead to random fluctuations of the angular position, indicating 
no explicit rotational motion of the triangle. This construct, consisting of a bead-spring triangle with force dipoles on each bond, is similar to the triangular micro-swimmer studied in Refs. \cite{swimmer_rizvi2018,rizvi_2020} in presence of hydrodynamic interactions. Notably, here we find that the Rouse dynamics (i.e., local friction), together with anisotropy of force dipole strength and random-telegraph fluctuations, is enough to produce rotational motion.}
\label{fig_triangle}
\end{figure}

\subsection{Network collapse: dynamical persistence length in a harmonic potential}
\label{append_sec_persistence}
The notion of ``dynamical persistence length" emerged in active systems, in analogy to the persistence length of polymers, to describe the motion under forces that exhibit temporal correlations \cite{dynamical_persistence_RMP_2016}. In the case of stochastic force that follows the random telegraph process, the correlation time $\tau$ can be thought of as ``persistent time" for which the force remains in an `on' state with magnitude $f_0$ and the velocity is kept persistent. For a free particle moving under such active force, one can define a dynamical persistence length, $x_p$, as the distance traveled in a time $\tau$, 
thereby $x_p=f_0\tau/\gamma$ \cite{dynamical_persistence_cao_2019,dynamical_persistence_RMP_2016}.

In the case of a particle bound to a harmonic potential, we generalize the above notion by using the equation of motion
\beq
\gamma\frac{dx}{dt}= -m\omega_0^2 x + f_0
\eneq
with $x(0)=0$, whose solution is
\beq
x(t) = \frac{f_0}{m\omega_0^2}\left(1-e^{-t/\tau_0}\right),
\eneq
where $\tau_0=\gamma/m\omega_0^2$. Using $t=\tau$, we obtain the desired persistence length
\beq
x_p = \frac{f_0}{m\omega_0^2}\left(1-e^{-\tau/\tau_0}\right)\;,
\eneq
which (obviously) coincides with the free-particle definition for $\tau\ll \tau_0$. The critical force above which one\
 expects a collapse of the network is obtained by equating $x_p=b$, leading to
\beq
f_{0c}\simeq \frac{m\omega_0^2 b}{\left(1-e^{-\tau/\tau_0}\right)}.
\eneq
This reduces to the familiar critical force  $f_{0c}\simeq \gamma b/\tau$ for $\tau\ll\tau_0$, and converges to $f_{0c\
}\simeq m\omega_0^2 b$ for $\tau\gg\tau_0$.

\section{Viscolelastic solvent}\label{append_sec_viscoelastic} 
While our focus in this paper has been on a viscous solvent, it is interesting to evaluate the asymptotic anomalous subdiffusion, in the case of force monopoles, to viscoelastic fluid. This was studied for a thermal system of linear polymer and applied to chromatin modeled as a space-filling curve \cite{viscoelastic_weber_2010,visco_vandebroek_2015,viscoelastic_polovnikov2018}. For a ``power-law solvent", that is for a solvent whose complex modulus follows a power-law, $G^*(\omega)\sim (i\omega)^{\alpha}$, it was shown that the thermal subdiffusion exponent is modified from $\nu_{th}=1/2$, the well-known Rouse subdiffusion exponent (that is associated in the present formalism with $d_s=1$) to $\nu_{th}=\alpha/2$. Hence, if one assumes $\alpha=0.8$, one obtains $\nu_{th}=0.4$, that was experimentally observed. However, for an active linear polymer in a viscoelastic solvent, it was shown \cite{visco_vandebroek_2015} that $\nu_{ac}=3\alpha/2-1$ (omitting the contribution of the CM motion). Thus, for $\alpha=0.8$, one obtains $\nu_{ac}=0.2$. Hence, a viscoelastic power-law solvent cannot explain the experimentally observed equality between the passive and active exponents, $\nu_{ac}=\nu_{th}=0.4$.

Here, we generalize these results for an arbitrary fractal. We do not provide here a complete formal analytical derivation, as this is very lengthy and out of the scope of the present publication. Rather, we suggest a compelling argument for the above-stated result, which stems from similar works \cite{viscoelastic_weber_2010,viscoelastic_vandebroek2014,visco_vandebroek_2015,viscoelastic_polovnikov2018,viscoelastic_granek2011,mckintosh_2021}.

Consider, first, the thermal MSD. For a viscous solvent the bead friction coefficient $\gamma$ is proportional to the solvent viscosity $\eta$, and from Eqs.\ (\ref{append_thermal_msd})-(\ref{append_thermal_msd_Bth}) (with $\nu_{\text{th}}=1-\frac{d_s}{2}$) we have
\beq
\langle\Delta\vec{r}(t)^2\rangle_{\text{th}}\sim \eta^{-(1-\frac{d_s}{2})}t^{1-\frac{d_s}{2}}
\eneq
Consider the MSD evaluated in the Fourier-Laplace plane (denoted by $t\to i\omega$) for a viscoelastic fluid. Replacing $\eta$ by the complex frequency-dependent viscosity $\eta^*(\omega)$, we have
\beq
\langle\Delta\vec{r}(i\omega)^2\rangle_{\text{th}}\sim \eta^*(\omega)^{-(1-\frac{d_s}{2})}(i\omega)^{-(2-\frac{d_s}{2})}
\eneq
Using the identity $G^*(\omega)=i\omega\eta^*(\omega)$, we have
\beq
\langle\Delta\vec{r}(i\omega)^2\rangle_{\text{th}}\sim G^*(\omega)^{-(1-\frac{d_s}{2})}(i\omega)^{-1}
\eneq
and assuming $G^*(\omega)\sim (i\omega)^{\alpha}$ we obtain
\beq
\langle\Delta\vec{r}(i\omega)^2\rangle_{\text{th}}\sim (i\omega)^{-\alpha(1-\frac{d_s}{2})-1}
\eneq
Transforming back to the time domain we find
\beq
\langle\Delta\vec{r}(t)^2\rangle_{\text{th}}\sim t^{\alpha(1-\frac{d_s}{2})}
\eneq
such that
\beq
\nu_{th}=\alpha\left(1-\frac{d_s}{2} \right)
\eneq
\begin{figure}
\centering
\includegraphics[width=0.41\textwidth]{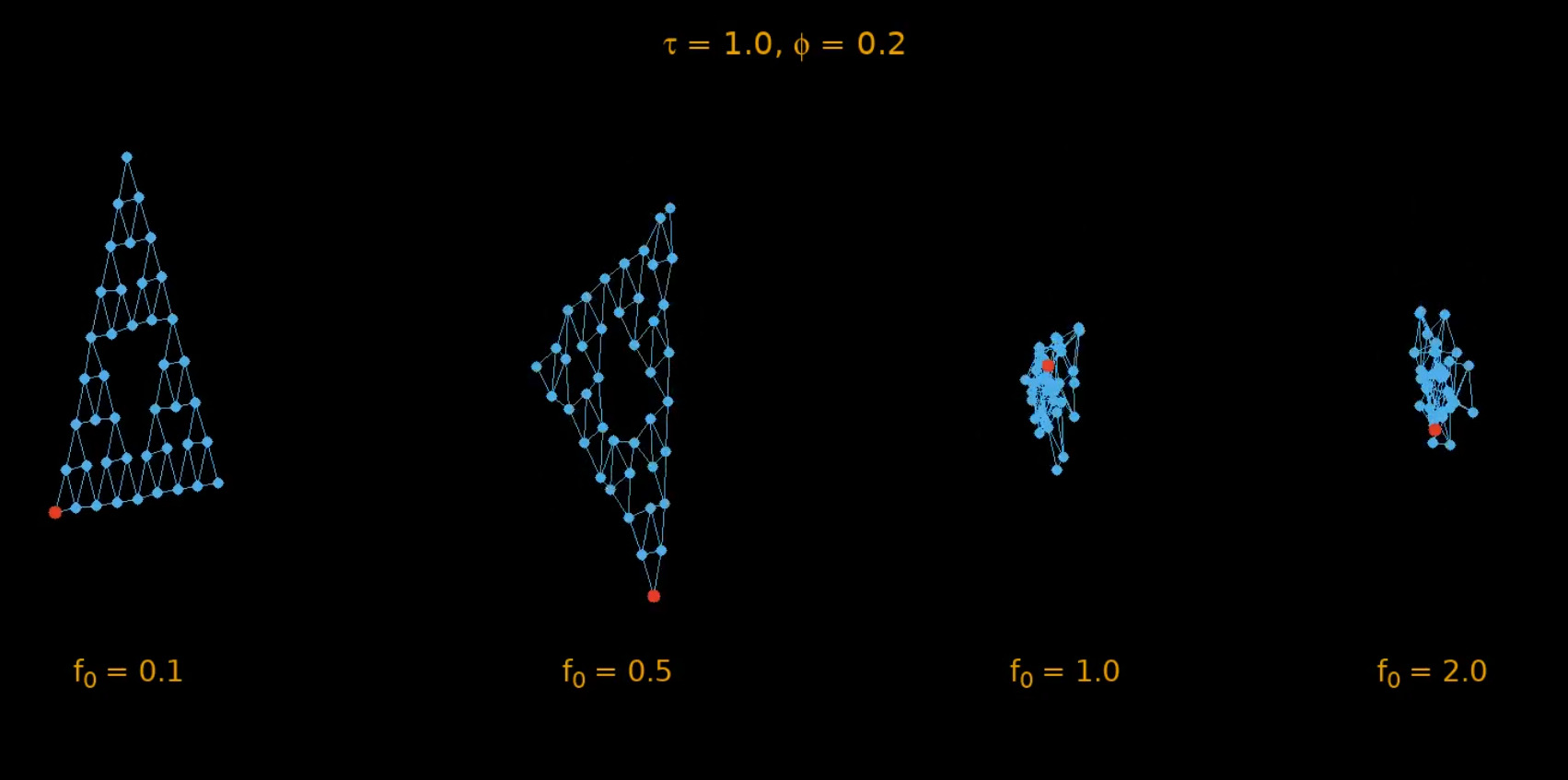}
\caption{A snapshot of the video (fig15.mp4) showing the motion of Sierpinski gaskets for different force amplitudes $f_0$ of force dipoles (Multimedia available online).}
 \label{sm1}
\end{figure}
\begin{figure}
\centering
\includegraphics[width=0.41\textwidth]{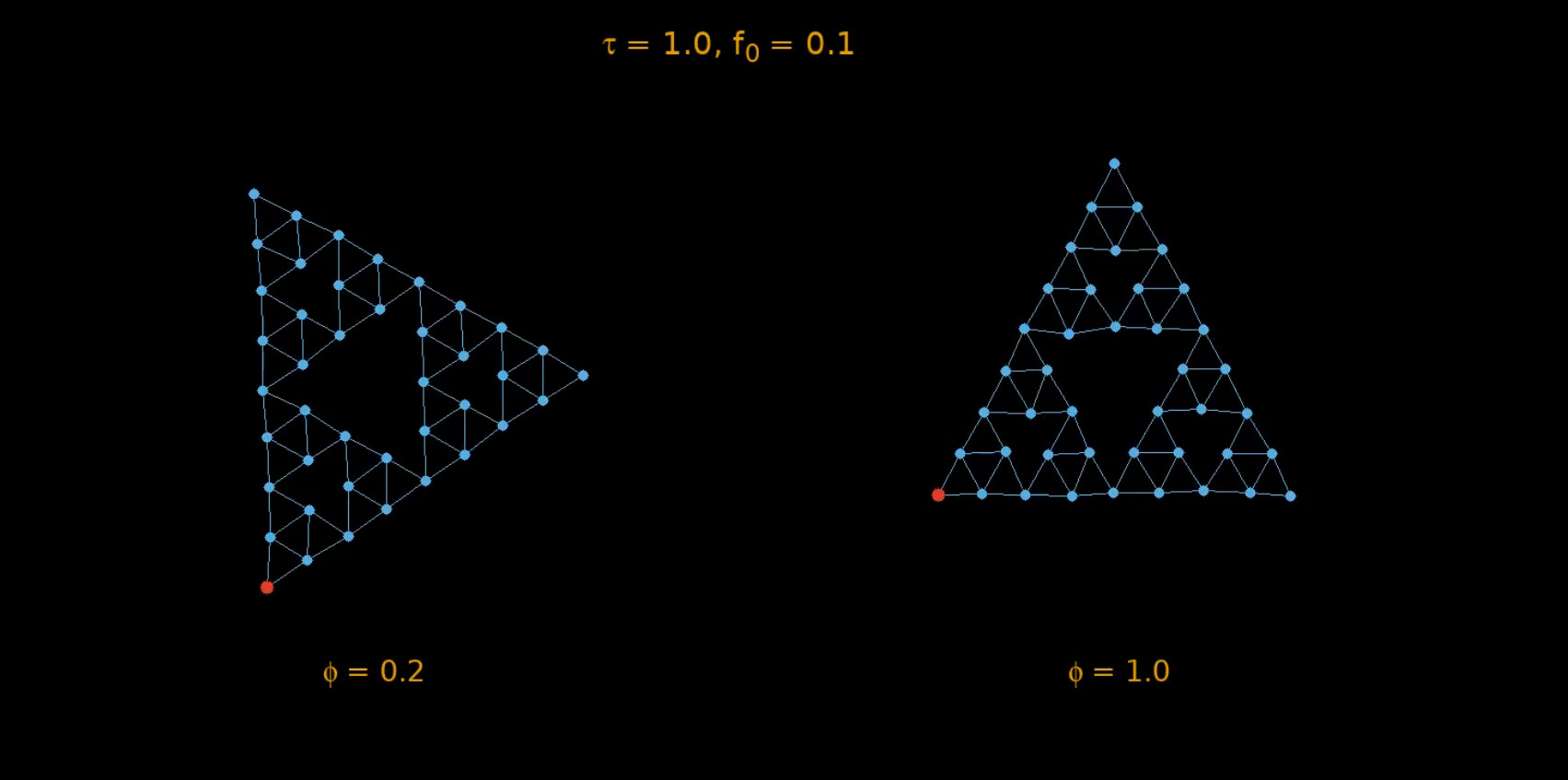}
\caption{A snapshot of the video (fig16.mp4) showing the motion of Sierpinski gaskets under different fraction $\phi$ of force dipoles (multimedia available online).}
 \label{sm2}
\end{figure}
\begin{figure}
\centering
\includegraphics[width=0.41\textwidth]{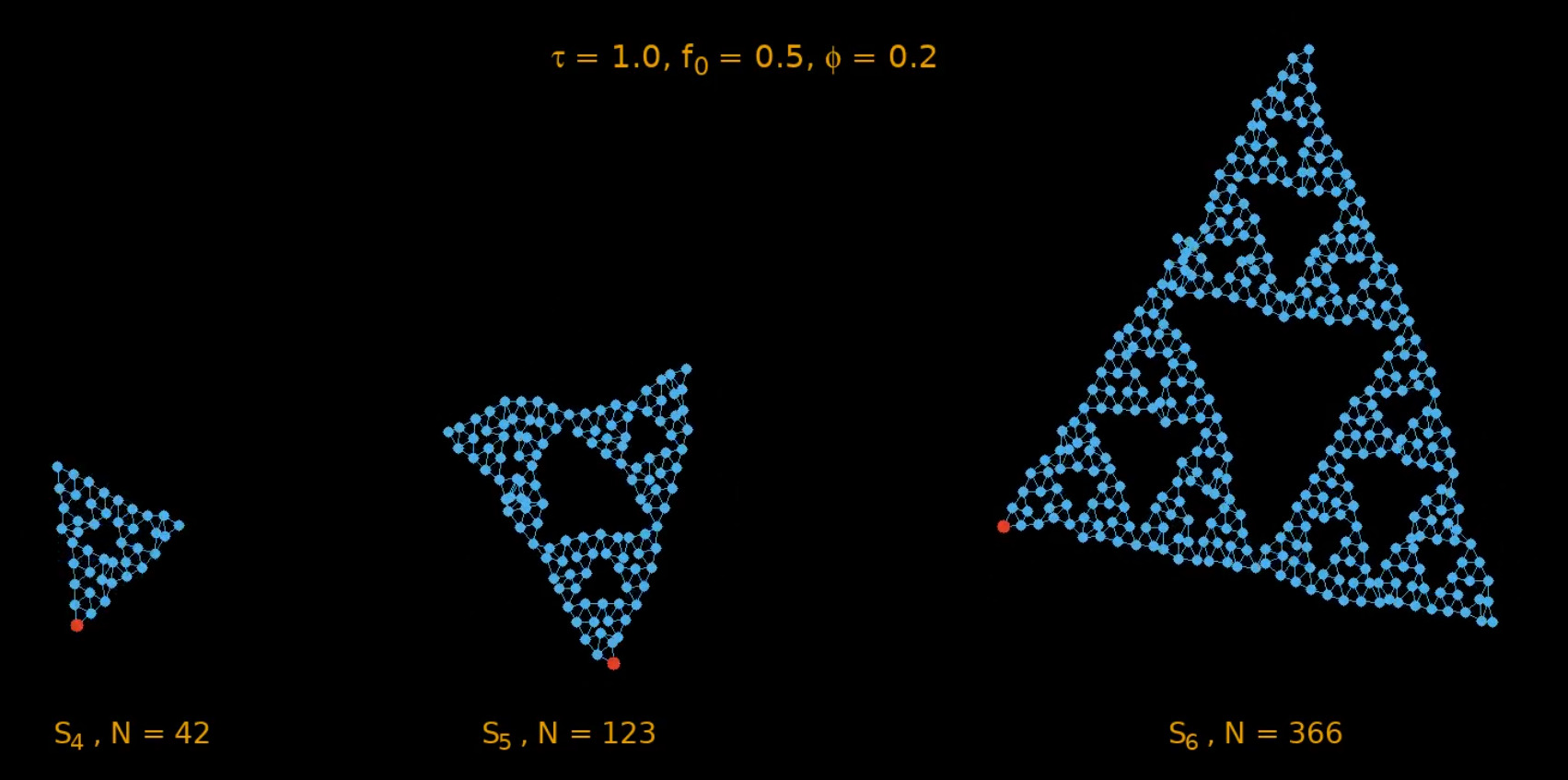}
\caption{A snapshot of the video (fig17.mp4) showing the motion of different generations of gaskets under force dipoles (multimedia available online).}
 \label{sm3}
\end{figure}

Next, consider the active network bead MSD in a viscoelastic solvent. Since for the viscous solvent case
\beq
\langle\Delta\vec{r}(t)^2\rangle_{\text{th}}\sim \eta^{-(2-\frac{d_s}{2})}t^{1-\frac{d_s}{2}}
\eneq
it follows that for the case of a viscoelastic solvent
\beq
\langle\Delta\vec{r}(i\omega)^2\rangle_{\text{th}}\sim \eta^*(\omega)^{-(2-\frac{d_s}{2})}(i\omega)^{-(2-\frac{d_s}{2})}
\eneq
leading to
\beq
\langle\Delta\vec{r}(i\omega)^2\rangle_{\text{th}}\sim G^*(\omega)^{-(2-\frac{d_s}{2})}
\eneq
and with $G^*(\omega)\sim (i\omega)^{\alpha}$ we obtain
\beq
\langle\Delta\vec{r}(i\omega)^2\rangle_{\text{th}}\sim (i\omega)^{-\alpha(2-\frac{d_s}{2})}
\eneq
Transforming back to the time domain, we find
\beq
\langle\Delta\vec{r}(t)^2\rangle_{\text{th}}\sim t^{\alpha(2-\frac{d_s}{2})-1}
\eneq
i.e.,
\beq
\nu_{ac}=\alpha\left(2-\frac{d_s}{2} \right)-1
\eneq
These exponents agree with the above-cited linear polymer case, $d_s=1$.

Thus $\nu_{ac}\neq \nu_{th}$ unless $\alpha=1$, i.e., equality is recovered only for the case of a viscous fluid studied in this paper. Hence, whatever the value of $d_s$, a power-law embedding viscoelastic fluid cannot explain equality between the passive and active subdiffusion exponent.

\bibliography{active}

\end{document}